\documentclass[
  aps,
  pre,
  reprint,
  superscriptaddress,
  longbibliography
]{revtex4-2}

\usepackage{graphicx}
\usepackage{amsmath,amsfonts,amssymb}
\usepackage[colorlinks=true,
            linkcolor=blue,
            citecolor=blue,
            urlcolor=blue]{hyperref}

\def\beq{\begin{equation}\begin{aligned}}
\def\eeq{\end{aligned}\end{equation}}

\begin{document}

\title{Stabilizing Role of Uninformed Participants in Collective Decision Making}

\author{Leonardo Colombo}
\affiliation{Centre for Automation and Robotics (CSIC-UPM), Madrid, Spain}

\author{Mar\'ia Emma Eyrea Iraz\'u}
\affiliation{Departamento de Matem\'atica, Centro de Matem\'atica (CMaLP), UNLP, Argentina}

\author{Laura P.\ Schaposnik}
\affiliation{Department of Mathematics, Statistics and Computer Science, University of Illinois Chicago, Chicago, IL, USA}
\affiliation{NSF-Simons National Institute for Theory and Mathematics in Biology, Chicago, IL, USA}
\affiliation{The Laboratory for Quantum Theory at the Extremes (LQuTE),  IL, USA}
\author{James Unwin}

\affiliation{The Laboratory for Quantum Theory at the Extremes (LQuTE),  IL, USA}
\affiliation{Department of Physics, University of Illinois Chicago, Chicago, IL, USA}
\date{\today}

\begin{abstract}
For groups without strict hierarchy, collective decisions often emerge through compromise. We develop a second-order network model of collective decision-making using a dissipative Hamiltonian formulation, in which informed agents introduce preferred directions while uninformed participants contribute only direction-free dissipation. We show that under low conflict, the model admits a locally unique, exponentially stable compromise state. Using a structured modular network we further show that as conflict increases the local compromise branch terminates through a saddle-node fold rather than through a smooth mean-field symmetry-breaking transition. Modular polarized states persist on branches that are locally separated from
the compromise branch. Direction-free dissipation does not shift the static structural threshold, but it delays escape from the saddle-node ghost and pushes the observable onset of polarization to larger conflicts. Our work  identifies a dissipation-mediated mechanism, complementary to connectivity-based accounts, through which uninformed participants stabilize collective behavior in biological and engineered swarms.
\end{abstract}

\maketitle


\section{Introduction}

Collective decisions in animal groups, human crowds, and robot swarms often emerge from the interplay between a small minority of informed leaders and a larger population of uninformed followers. Classic studies in schooling fish, flocking birds, and social insects show that groups may either select one of several competing options or settle on a \emph{compromise} direction~\citep{couzin2005effective,leonard2012decision,herbert2011inferring,krause2002living,conradt2005consensus,sumpter2008consensus}. Experiments and theory suggest that uninformed individuals can strongly reshape these outcomes \cite{couzin2011uninformed}. Existing explanations have emphasized connectivity, local sensing, or population-balance effects~\citep{couzin2005effective,leonard2012decision,herbert2011inferring}. Complementary to these perspectives, here we argue that uninformed individuals can stabilize collective compromise.

The central idea is that direction-free dissipation stabilizes compromise by acting on the slow modes that first destabilize agreement. We study this within a second-order network model. In our formulation, uninformed individuals contribute no decision-relevant bias, but do contribute damping along the weakest modes of the interaction graph. This mechanism delays the observable onset of polarization, prolongs traversal of the near-rupture bottleneck, and enlarges the functional regime over which the group appears to maintain compromise.

Over the past two decades, numerous models have sought to explain how collective choices emerge under competing preferences. Vicsek-type models~\cite{vicsek1995novel} capture alignment and coherence, but provide limited insight into multi-option decision-making. Leader-follower frameworks~\cite{couzin2005effective,leonard2012decision,yates2009inherent} quantify how informed individuals steer the group and how uninformed individuals reshape outcomes through connectivity and sensing effects, but they do not isolate a dissipation-mediated mechanism for stabilizing compromise. Majority-vote and Bayesian aggregation models~\cite{arganda2012common,marshall2009optimal} can predict collective choices, yet typically do not provide a dynamical account of transitions between compromise, polarized decision, and fragmentation.

Across these frameworks, two limitations are particularly relevant here. First, many models do not include an explicit second-order mechanism by which individuals accumulate and dissipate internal activity or commitment over time, despite strong empirical support for such processes in animal and human decision-making~\citep{marshall2009optimal,deneubourg1990self,gold2007neural,ratcliff2008diffusion}. Second, existing approaches do not explicitly connect agent-level dissipation and network structure to the loss of compromise, the emergence of disconnected decision branches, and the shift between static structural limits and finite-time observable transitions. Prior studies have explored the implications of uninformed individuals within connectivity- and sensing-mediated models~\citep{leonard2012decision}.

Here we take a different approach, proposing a dissipation-mediated second-order dynamical model in which uninformed participants play a stabilizing role in collective decision-making. Agents interact through power-conserving couplings, while intrinsic damping removes internal activity without introducing directional bias. An auxiliary internal coordinate tracks the cumulative balance between injected and dissipated activity. At the agent-level with observable variables \((\theta_i,p_i)\), the dynamics form a damped second-order Kuramoto-type system \cite{kuramoto1975international}. Moreover, adopting a port-contact-Hamiltonian formulation makes dissipation intrinsic to the geometry and yields a transparent network-level energy-dissipation identity through which collective interactions can redistribute activity but cannot create it.

We establish three main results. First, for low conflict and sufficient connectivity, the group admits a unique, exponentially stable compromise equilibrium. Second, in the structured modular networks considered here, increasing conflict need not lead to loss of compromise through a smooth mean-field symmetry-breaking transition. Instead, the compromise branch can terminate through a saddle-node fold associated with bottleneck-limited transmission of alignment,  while polarized decision states exist on branches that do not emerge locally from the compromise state. Third, finite-time dynamics near rupture are governed by delayed escape via the saddle-node ghost, producing an observable transition that is shifted away from the static structural threshold.

Uninformed individuals play a crucial role in this mechanism. By contributing dissipation without directional bias, they damp the slow collective modes that control the approach to rupture and thereby delay the observable onset of polarization. In biological and engineered collectives alike, uninformed agents enhance cohesion not by providing information, but by reshaping the decision landscape of the system.
\smallbreak


\section{Model and energy-dissipation structure} 

We start by introducing the dynamics, the interaction network that couples agents, and an energy-dissipation viewpoint that clarifies the stabilizing role of direction-free damping. This will also allow us to distinguish between smooth mean-field transitions and topology-induced rupture mechanisms in structured networks.



\subsection{Structure}
We shall represent the group as a collection of \(N\) interacting agents, each carrying three state variables that describe how individuals adjust their motion and respond to social and directional cues. The heading \(\theta_i\in\mathbb{S}^1\) is the observable decision variable, indicating the direction of motion or the option currently expressed by the agent.
The subscript \(i\in\{1,\dots,N\}\) indexes agents.
The momentum-like variable \(p_i\in\mathbb{R}\) determines how rapidly the heading changes and captures behavioral inertia (response latency) to new inputs.

The internal variable $z_i \in \mathbb{R}$ is a contact-geometric bookkeeping coordinate that tracks the cumulative balance between injected and dissipated activity associated with alignment and leader forcing. In the absence of sustained interactions, this quantity decays over time. Its role is not to introduce an additional observable decision variable, but to make dissipation intrinsic to the agent-level geometry and thereby enable a closed energy-dissipation identity at the network level. 
The triplet $(\theta_i,p_i,z_i)$ therefore captures the expressed decision, response latency, and cumulative dissipative balance at the agent level.

The model is formulated as a network of \emph{port-contact-Hamiltonian systems}, explicitly
\begin{equation}
\begin{aligned}
K_i
&=
\Big(\frac12 p_i^2-\alpha_i\cos(\theta_i-\phi_i)\Big)\\[8pt]
& + 
\Big(-\kappa\sum_j a_{ij}\cos(\theta_j-\theta_i)\Big)
 + 
\gamma_i z_i.
\label{eq:contact-Hamiltonian}
\end{aligned}
\end{equation}
Importantly, the coordinate value $z_i$ does not explicitly modulate the forces in the
$(\theta_i,p_i)$-subsystem; rather, including the term $+\gamma_i z_i$ in the contact Hamiltonian
ensures $\partial_{z_i}K_i=\gamma_i$, which structurally generates the momentum damping term
$-\gamma_i p_i$ in the contact equations of motion (via the $-p_i \partial_{z_i}K_i$ contribution).
See SI for the associated equations of motion (EoM).

The first bracketed term is the local conservative storage, the next bracketed term provides power-conserving alignment interactions across the network, and the final term is the contact contribution through which dissipation is represented intrinsically. Together, these ingredients define a geometric framework that combines interaction, irreversible loss, and network-level passivity.

The coefficients \(a_{ij}\ge 0\) define the weighted adjacency matrix \(A\in\mathbb{R}^{N\times N}\), with \(a_{ij}=0\) for non-neighboring agents. The constant \(\kappa>0\) is the \emph{alignment gain}, which sets the strength of social coupling, determining how strongly agents act to reduce heading differences with their neighbors. Specifically, increasing \(\kappa\) accelerates the relaxation of local disagreements and enhances the effectiveness of network-mediated coordination. 
The parameters \(\alpha_i\ge 0\) encode the strength of directional preference towards a target direction \(\phi_i\), while \(\gamma_i>0\) represents local damping. We use \emph{uninformed participants} for agents with negligible directional preference ($\alpha_i\simeq 0$),
and we refer to the resulting stabilization mechanism as \emph{direction-free dissipation}.

Notably, analogies can be drawn between the structure above and certain physical systems. The network coupling studied here is reminiscent of a damped spring network and frustrated quantum/classical  many-body systems (see e.g.~\cite{grason2016perspective,Giampaolo:2011fqf}). Notably, these systems admit similar Hamiltonian descriptions and exhibit competition between incompatible local preferences and global compatibility constraints. Parallels between physical systems and collective decision-making models, and the potential to provide interesting insights.


\subsection{Network}
Information flow through the group is shaped by the interaction network.
For a symmetric adjacency matrix \(A\), the graph Laplacian is \(L=D-A\), where
\(D_{ii}=\sum_{j=1}^{N} a_{ij}\).
The Laplacian governs how heading differences are redistributed through alignment interactions.
Its second-smallest eigenvalue \(\lambda_2\) (the Fiedler eigenvalue or algebraic connectivity), measures the weakest collective coordination mode.
Large \(\lambda_2\) implies fast decay of non-consensus perturbations, whereas small \(\lambda_2\) allows slow modes that can support polarization or fragmentation.

Informed leaders are characterized by $\alpha_i>0$ and preferred directions $\phi_i$, while uninformed individuals have $\alpha_i \approx 0$. Both classes may have $\gamma_i>0$, but uninformed agents contribute no directional preference and primarily affect the dynamics through damping. 
The mode associated to the eigenvalue $\lambda_2$ remains central in the low-conflict compromise regime, while in structured networks its interaction with topological bottlenecks will later help determine how compromise is lost. Specifically, when coordination must pass through sparse inter-modular cut-sets or bridges (i.e.~a small set of edges whose removal would disconnect the modules), these topological bottlenecks can limit the transmission of alignment across the graph and thereby alter the mechanism through which compromise is lost. Despite these topological limitations, uninformed damping suppresses growth along the available slow collective modes and enlarges the parameter region in which collective compromise remains stable.

A central property of the model is that alignment forces arise through a power-conserving interconnection among agents, thus interactions can redistribute energy across the network, but they cannot create it.
As a result, the only truly irreversible losses arise from the local damping terms \(\gamma_i\), which permanently remove energy and represent the gradual decay of momentum at the individual level.

We distinguish the contact Hamiltonian $K(\theta,p,z)$, which generates the dissipative dynamics, from the stored energy $H(\theta,p)$, which quantifies passivity at the network level. In our formulation, $K$ depends explicitly on $z$, and this additional coordinate provides the contact-geometric embedding through which dissipation is represented intrinsically. By contrast, the stored energy $H$ is associated only with the conservative part of the dynamics and does not depend on $z$. The total stored energy associated with the conservative part of the dynamics is given by

\begin{equation}
\begin{aligned}\notag
H=
\sum_{i=1}^N \Big(\tfrac12 p_i^2 - \alpha_i \cos(\theta_i-\phi_i)\Big)
-\frac{\kappa}{2}\sum_{i,j} a_{ij}\cos(\theta_j-\theta_i).
\end{aligned}
\end{equation}
The evolution of $H$ along trajectories defined by the EoM 
satisfies the energy-dissipation identity (derived in SI) 
$\dot H
=  - \sum_{i=1}^N \gamma_i p_i^2
 \le  0$
which shows that alignment interactions are power-conserving, whereas local damping removes energy irreversibly and suppresses collective instabilities.
Thus, the total stored energy can never increase. The network exhibits a passive energy balance, since internal interactions do not generate energy, and all amplification mechanisms must compete with dissipation. This global constraint explains why dissipation suppresses the growth of instabilities and helps determine whether conflict is absorbed into compromise, delayed dynamically, or redirected into topology-dependent rupture scenarios in structured networks.

\subsection{Macroscopic observables}
\label{2.3}

To characterize collective behavior, we formulate macroscopic observables that summarize the microscopic configuration. We introduce the complex order parameter
\begin{equation}
\label{rpsi}
r e^{\mathrm{i}\psi} = \frac{1}{N}\sum_{i=1}^N e^{\mathrm{i}\theta_i}
\end{equation}
The magnitude \(r\in[0,1]\) tracks the degree of global alignment. Large \(r\) indicates a cohesive group, whereas small \(r\) signals weak global coordination. We distinguish three macroscopic regimes: compromise, in which \(r\approx 1\) and \(|L_A-L_B|\approx 0\); polarized decision, in which \(r\approx 1\) and \(|L_A-L_B|\) is large; and fragmentation, in which global order is low while module-level order remains high, so that distinct subgroups remain internally coherent but fail to align with one another globally. The average group heading \(\psi\) is defined as the phase of the mean complex vector.

When two competing preferred directions \(\phi_A\) and \(\phi_B\) are present we denote their angular separation by the {\em conflict angle} $\Delta := |\phi_A-\phi_B| \in [0,\pi]$. We then quantify the group's bias toward each option through ($\omega=A,B$)
\[
L_\omega = \frac{1}{N}\sum_{i=1}^N \cos(\theta_i-\phi_\omega),
\]
We emphasize $\theta_i$ is the instantaneous heading, whereas $\phi_i$ is the preferred heading associated to a directional bias. 
The difference \(L_A-L_B\) provides a macroscopic collective decision bias, and $\Delta$ quantifies disagreement between leader-preferred directions. In addition to such global observables, structured networks may require edge- or cut-set-based quantities to diagnose how compromise is lost. In particular, when the interaction graph contains sparse bridges or bottlenecks, the inter-module phase tension can become a more informative indicator of impending rupture than the global decision bias alone.

\begin{figure}[t]
   \centering
\includegraphics[width=\linewidth]{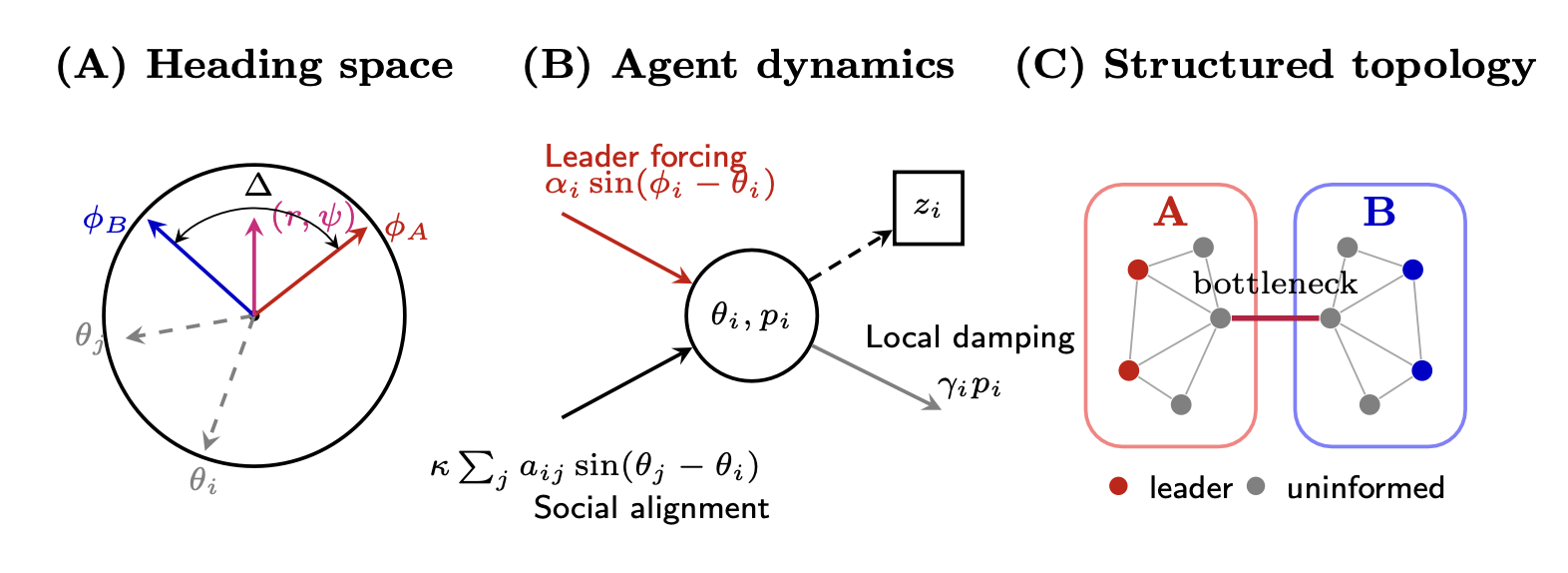}
\vspace{-2mm}  
\caption{\textbf{Schematic of the model.}
(A) Heading space representing individual states $\theta_i$, the order parameter $(r,\psi)$, and the conflict angle $\Delta$ between leader-preferred directions $\phi_A$ and $\phi_B$.
(B) Single-agent dynamics driven by leader forcing, social alignment, and local damping $\gamma_i$, with auxiliary coordinate $z_i$.
(C) Structured interaction graph with two dense modules connected through a sparse bridge, forming a topological bottleneck.}
\label{fig:model-schematic}
\end{figure}

Figure~\ref{fig:model-schematic} summarizes the relation between agent-level dynamics, network-mediated alignment, and the macroscopic observables used to distinguish compromise, polarized decision, fragmentation, and bottleneck-mediated rupture. It also highlights how structured interaction topology constrains the transmission of coordination across the group.

\section{Main results}

In this section, we characterize the collective dynamics across varying levels of conflict \(\Delta\) between two competing target directions \(\phi_A\) and \(\phi_B\). We first demonstrate that when leader disagreement is weak, the network naturally sustains a stable collective compromise. We then show how, in the modular regime studied here, this compromise can be lost through bottleneck-mediated rupture, producing separated polarized states and, in the numerics, bistability and hysteresis.

\subsection{Stable compromise under low conflict}
\label{3.1}
The case of low conflict corresponds to $\Delta\ll1$.
The two leaders have an average preferred direction
\[
\bar{\phi}:=\hbox{Arg} \left(\alpha_A e^{\mathrm{i}\phi_A}+\alpha_B e^{\mathrm{i}\phi_B}\right),
\]
and this provides a natural reference frame in this case.
Working in relative coordinates \(\tilde{\theta}_i=\theta_i-\bar{\phi}\), we analyze the dynamics on an undirected connected interaction graph with algebraic connectivity \(\lambda_2>0\). The quantity \(\lambda_2\) sets the scale of the weakest non-consensus coordination mode, and therefore of the slowest collective deformation through which disagreement relaxes.

Let \(v_2\) denote the Laplacian eigenvector associated with \(\lambda_2\), i.e., the Fiedler mode. This mode identifies the weakest coordination direction of the network and will organize the low-conflict relaxation mechanism below.

\vspace{-2mm}
\subsection*{Theorem 1 (Stable compromise)}
\emph{Assume the interaction graph is undirected and connected, and that at least one agent is informed, i.e.\ $\alpha_i>0$ for some $i$. If the conflict $\Delta$ is sufficiently small, then the system admits a unique compromise equilibrium $(\tilde{\boldsymbol\theta}^\ast(\Delta),\boldsymbol p^\ast)$ near $(0,0)$, with $\boldsymbol p^\ast=0$. Further, this equilibrium is locally exponentially stable.}

\vspace{1mm}
The proof is given in the SI. This result formalizes the intuitive idea that weak conflict promotes collective compromise.
The `small' regime of $\Delta$ is characterised by the size of $\kappa\lambda_2$, and thus stronger coupling enhances the valid range for which $\Delta$ may be taken as small.
Notably, this regime is organized by damped second-order network modes. The Fiedler direction determines the weakest relaxation channel, while direction-free dissipation from uninformed individuals strengthens its contraction without favoring either option. Together, these ingredients create a parameter region in which groups reliably converge to a shared direction before polarization emerges.

\begin{figure*}[t!]
    \centering
    \includegraphics[width=\textwidth]{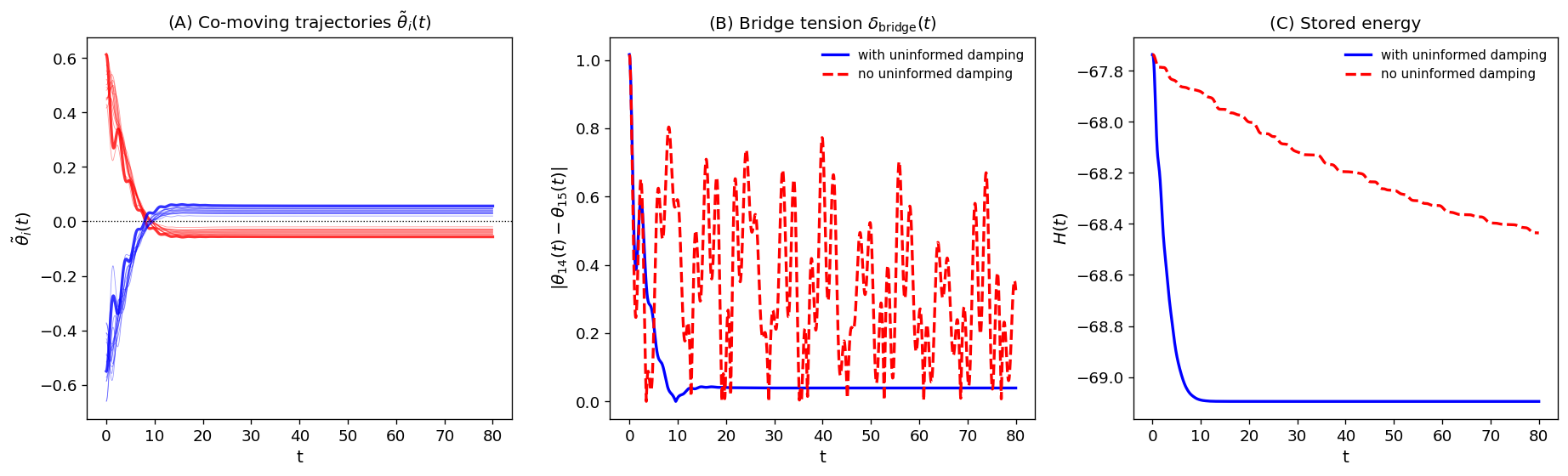}
   \caption{
Simulations of the stable-compromise regime. Dynamics of $N=30$ agents on a modular interaction network comprising two densely connected modules coupled by a single bridge ($\lambda_2 \approx 0.078$, $\kappa=1.10$, conflict $\Delta=0.3$). Leader damping is $\gamma_{\mathrm{leader}}=0.30$; uninformed damping is $\gamma_{\mathrm{uninformed}}=0.70$ (blue, solid) or $0$ (red, dashed). The system is initialized with a moderate modular bias to highlight transient relaxation. (A) Individual trajectories $\tilde{\theta}_i(t)$ in the co-moving frame, colored by module (red vs.\ blue), showing convergence to a structured, cohesive compromise state. (B) Inter-module bridge tension \(\delta_{\mathrm{bridge}}(t)\).
Without uninformed damping, the tension exhibits larger oscillatory
transients, whereas uninformed damping rapidly suppresses these
fluctuations. (C) Overall decay of stored energy $H(t)$, consistent with the network's passive energy balance.} 
\label{fig:theorem1-simulations}
\end{figure*}

Fig.~\ref{fig:theorem1-simulations} illustrates the stable-compromise regime predicted by Theorem~1. The trajectories converge to a cohesive compromise state, the inter-module bridge tension remains bounded and is strongly suppressed by uninformed damping, and the stored energy decays monotonically in agreement with the network's passive energy balance. See SI for further discussion.

\subsection{Topological rupture and the collapse of compromise}

Since the compromise equilibrium identified in Section \ref{3.1} is stabilized by the network's internal restoring forces, its persistence depends critically on the capacity of the interaction graph to transmit alignment across the network. Historically, the emergence of collective decisions in polarized groups has been modeled as a continuous, symmetry-breaking phase transition. 
In mean-field or globally coupled reductions of animal group decision dynamics, loss of the symmetric compromise state is organized by a pitchfork-type symmetry-breaking bifurcation \cite{nabet2009dynamics}, within the broader decision-versus-compromise framework of \cite{leonard2012decision}.

However, biological and engineered collectives operate on sparse, structured networks characterized by topological bottlenecks. In such networks, a smooth symmetry-breaking transition need not be the relevant mechanism for loss of compromise.     As leader disagreement increases, the compromise branch is increasingly constrained by the finite capacity of sparse network bottlenecks to transmit alignment. 
Because the pairwise alignment force is bounded and is maximized when phase differences approach \(\pi/2\), sparse inter-modular bottlenecks can constrain how much conflict the network can absorb before the compromise branch loses viability. To formalize this, we analyze the stability of the compromise state in a modular network with sparse inter-module coupling as the conflict $\Delta$ increases.


We now specialize to a modular network consisting of two connected modules, \(M_A\) and \(M_B\), coupled by a sparse cut-set \(C\). In mean-field settings, loss of compromise is often associated with a symmetry-breaking pitchfork. In the modular bottleneck regime considered here, by contrast, the compromise branch can terminate through a fold of the reduced two-module dynamics. For sufficiently small \(\Delta\), let \(\tilde{\boldsymbol\theta}^\ast(\Delta)\) denote the corresponding locally stable compromise branch.

For the modular network described above, we refer to
\[
\delta_{\mathrm{bridge}}^\ast(\Delta)
:=
\max_{(i,j)\in C}
|\tilde\theta_i^\ast-\tilde\theta_j^\ast|
\]
as the \emph{maximal bridge tension} of the compromise branch.
In the modular regime of interest, this quantity quantifies how disagreement is transmitted across the bottleneck.

In what follows we shall specialize to a modular interaction graph consisting of two connected subgraphs (modules) $M_A$ and $M_B$,
coupled by a sparse cut-set of edges $C$ (a “bridge” in the simplest case)   of cross-module pairs \((i,j)\) with \(i\in M_A\), \(j\in M_B\), each edge counted once, with leaders in \(M_s\) preferring target \(\phi_s\), \(s\in\{A,B\}\), and conflict angle \(\Delta=|\phi_A-\phi_B|\).
We refer to the regime of
\emph{weak inter-module coupling} when the total weight across $C$ is small compared to intra-module coupling,
so each module is internally coherent while inter-module coordination is bottlenecked by $C$.
We shall define 
$\alpha^s := \sum_{i\in M_s}\alpha_i$ and
\[K_C := \kappa\sum_{(i,j)\in C} a_{ij},
\qquad
\eta := \frac{\max(\alpha^A,\alpha^B,K_C)}{\kappa\min_s \lambda_2(L_s)},
\]
where $L_s$ denotes the isolated graph Laplacian of module $M_s$, obtained by restricting the
interaction graph to the nodes in $M_s$ and omitting all cut-set edges in $C$; since each module
is connected, $\lambda_2(L_s)>0$.
For \(\eta\ll1\), each module remains internally rigid to leading order.

\vspace{-4mm}
\subsection*{Theorem 2. (Bottleneck-mediated fold of compromise)}
\emph{
Within the above setting, suppose that as the  conflict angle \(\Delta\) varies in \((0,\pi)\), the corresponding rigid-module reduced system admits a compromise branch that undergoes a nondegenerate saddle-node bifurcation at some
\(
\Delta_{\mathrm{fold}}^{(0)}\in(0,\pi),
\)
 Further, assume that the reduced compromise branch stays in the regime where each module is on the restoring side of its leader forcing.
Then, for \(\eta\) sufficiently small, the compromise branch persists for \(\Delta\in[0,\Delta_{\mathrm{fold}})\) and terminates at \(\Delta=\Delta_{\mathrm{fold}}\in(0,\pi)\) through a saddle-node bifurcation. Moreover, the associated critical mode is concentrated on the bridge and has the form
\[
\varphi = c_A \mathbf{1}_A + c_B \mathbf{1}_B + O(\eta),
\qquad c_A c_B<0.
\]
}

\vspace{1mm}
The proof is given in the SI.
%
%
Theorem~2 identifies a topological rupture mechanism for loss of compromise in the modular regime. When intra-module coherence is strong, the dominant instability is not an internal deformation of either module but a bridge mode that drives the two modules against one another. In this sense, the saddle-node is bottleneck-mediated, since compromise fails because the sparse cut-set can no longer transmit the phase mismatch imposed by conflicting leaders.
The above theorem is a local bifurcation statement, it identifies fold-type termination of the compromise branch under the stated hypotheses, but does not by itself
determine the global arrangement of the polarized decision branches. The numerical continuation below suggests that, in the modular regime studied here, this local fold is embedded within a broader global branch structure.



We continue in the modular setting introduced above, the interaction graph consists of two connected modules \(M_A\) and \(M_B\), coupled by a cut-set \(C\), and all informed agents in \(M_A\) prefer the target \(\phi_A\), while all informed agents in \(M_B\) prefer the target \(\phi_B\), with \(\phi_A\neq\phi_B\). We say that the network is in the \emph{weak inter-module coupling regime} if there exists \(\varepsilon_0>0\) such that 
\[
a_{ij}\le \varepsilon
\qquad\text{for all }(i,j)\in C,
\qquad
0\le \varepsilon<\varepsilon_0 .
\]

\vspace{-2mm}
\subsection*{Theorem 3 (Local separation of modular polarized states)} 
\emph{In the weak
inter-module coupling regime, for each fixed conflict value
\(\Delta > 0\), there exists a  polarized equilibrium, locally exponentially stable, in which the modules remain aligned near their respective preferred targets. Moreover, if the hypotheses of Theorem 2 hold, then this modular polarized equilibrium does not arise through a local bifurcation from the compromise branch at \(\Delta = \Delta_{\rm fold}\).}

\medskip

\begin{figure*}[t!]
    \centering
\includegraphics[width=0.99\linewidth]{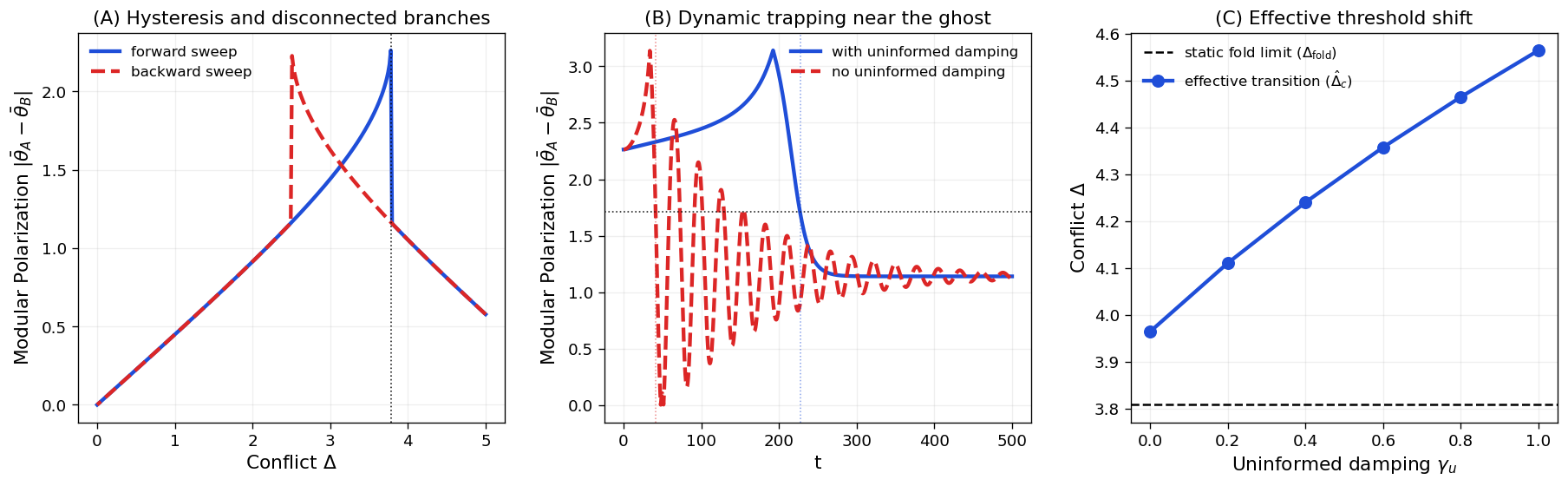}
\vspace{-2mm}  \caption{
Topological rupture and dynamical masking of collective breakdown in a modular network.
(A) Adiabatic forward and backward sweeps of the modular polarization \( |\bar{\theta}_A-\bar{\theta}_B| \) versus conflict \(\Delta\), revealing an abrupt jump and a hysteresis loop generated by the disconnected branches observed in numerical continuation.
(B) Dynamic trapping near the ghost of the saddle-node. Slightly beyond rupture, the undamped system (red, dashed) escapes quickly and oscillates strongly, whereas uninformed damping (blue, solid) produces a long-lived trapped transient near the remnant of the lost compromise state.
(C) Effective threshold shift under continuous conflict ramping. The observed transition \(\hat{\Delta}_c\) exceeds the static fold limit \(\Delta_{\mathrm{fold}}\) because of dynamic bifurcation delay and increases monotonically with uninformed damping. Passive participants therefore do not change the structural fold itself, but delay the finite-time onset of macroscopic rupture.
}
    \label{fig:theorem2-simulations}
\end{figure*}

Proofs are given in the SI. 
Theorem 3 does not fix the global branch geometry, rather, it establishes
the local existence and stability of a modular polarized state
that persists from the weakly coupled limit.
In the modular regimes studied numerically here, continuation
results further indicate that polarized branches extend away
from the local fold neighborhood and terminate at distinct
limit points, denoted schematically by \(\Delta_{\min}\).
 In the parameter line sweeps shown in Fig.~\ref{fig:theorem2-simulations}(A), the numerically observed ordering
\(\Delta_{\min}<\Delta_{\mathrm{fold}}\) yields a bistable interval in which both compromise and polarized decision states are present.
The broader multi-dimensional “hysteresis wedge” associated with this overlap is summarized schematically in Fig.~\ref{fig:global_geometry}B.

In finite networks and generic realizations, the transition via a smoothed symmetry-breaking bifurcation is typically not the most natural outcome. Rather, in the modular bottleneck regime studied here, the decisive event is the saddle-node loss of the compromise branch itself. The sudden, biased response then reflects a jump induced by branch disappearance, with subsequent relaxation toward one of the available polarized attracting states determined by the global phase-space geometry.

 To examine this branch geometry and test for the absence of a local pitchfork-type connection in the modular examples considered here, numerical continuation is useful. By tracking the steady states of the modular network as the conflict $\Delta$ is varied, we can explicitly visualize the fold termination of the compromise branch and the resulting geometric constraints that govern the macroscopic behavior. 

Theorem~2 makes a structural prediction that can be compared with the numerics of Fig.~\ref{fig:theorem2-simulations}. In Fig.~\ref{fig:theorem2-simulations}(A), the compromise branch terminates at a finite turning point \(\Delta_{\mathrm{fold}}\), rather than losing stability through a smooth mean-field-type onset, in agreement with the bottleneck-mediated fold described by the theorem. Together with the lower continuation limit \(\Delta_{\min}\) of the polarized branch, this produces a nonzero hysteresis interval \((\Delta_{\min},\Delta_{\mathrm{fold}})\), again consistent with the local fold loss of compromise and the separate existence of polarized states established by Theorems~2 and~3. Moreover, Fig.~\ref{fig:theorem2-simulations}(C) shows that the observed transition \(\hat{\Delta}_c\) satisfies \(\hat{\Delta}_c>\Delta{\mathrm{fold}}\), with the gap increasing under stronger uninformed damping, confirming that Theorem~2 identifies the static structural threshold while the visible macroscopic transition is shifted dynamically. Finally, the prominence of the inter-module bridge tension in Fig.~\ref{fig:theorem1-simulations}(B) and the modular contrast in Fig.~\ref{fig:theorem2-simulations}(A) are consistent with the Theorem 2 interpretation of the critical instability as an inter-module bottleneck mode rather than a bulk deformation of either module.

\subsection{Dynamic Delay of Polarization and Critical Slowing Down}
\label{3.3}

While Theorems 2 and 3 clarify the local static structure near loss of compromise and the existence of a stable modular polarized state in the weak inter-module coupling regime, biological and engineered collectives operate in real, finite time. Therefore, the observable emergence of macroscopic rupture is governed not only by the existence of topological bottlenecks, but also by the energy-dissipation balance that controls how the system dynamically traverses the post-fold phase space. Throughout, we focus on a fixed interaction graph, corresponding to a separation of timescales in which the macroscopic escape/delay dynamics occur faster than substantial network rewiring due to agent motion.

As the conflict approaches the structural limit ($\Delta \to \Delta_{\mathrm{fold}}$), the restoring force of the inter-modular bridge approaches its maximal capacity, and the Jacobian of the compromise branch approaches singularity. Just beyond the bifurcation ($\Delta \gtrsim \Delta_{\mathrm{fold}}$), the compromise equilibrium no longer exists. However, because the phase space is strongly pinched in this regime, trajectories do not depart immediately. Instead, they remain temporarily constrained near the remnant of the destroyed equilibrium, a bottleneck in the vector field often referred to as the \emph{ghost} of the saddle-node. This produces a slowing of the dynamics before the system relaxes toward a polarized branch selected by the global phase-space geometry.

Because uninformed agents ($\alpha_i \approx 0$) contribute direction-free dissipation ($\gamma_i > 0$) directly to the slow longitudinal deformation modes of the network, they act as viscous dampers exactly where the restoring and driving forces nearly balance. Consequently, increasing the fraction or damping coefficient of uninformed agents decreases the phase velocity through the bottleneck and substantially prolongs the time spent near the ghost.

Figure~\ref{fig:theorem2-simulations}(B) provides a direct numerical illustration of this mechanism. For a fixed conflict slightly beyond the rupture point, the undamped collective escapes rapidly from the ghost and undergoes a large oscillatory transient toward the disconnected branch. By contrast, uninformed damping strongly delays this escape, producing a long-lived trapped transient near the remnant of the lost compromise state before the eventual fall.

This delay has an important observational consequence. In a static bifurcation diagram, the structural limit is fixed by the fold at $\Delta_{\mathrm{fold}}$. Under finite-time observations or slowly varying environments, however, the transition is recorded only when the trajectory has actually escaped the ghost. As shown in Figure~\ref{fig:theorem2-simulations}(C), when the conflict is escalated continuously with time, the effectively observed threshold $\hat{\Delta}_c$ lies above $\Delta_{\mathrm{fold}}$ even in the absence of uninformed damping, reflecting a genuine dynamic bifurcation delay. Increasing uninformed damping shifts $\hat{\Delta}_c$ still further to larger conflict values.
Crucially, uninformed individuals do not alter the static topological capacity of the network, and the geometric fold remains fixed at $\Delta_{\mathrm{fold}}$. What they modify is the dynamical time scale on which the catastrophic transition becomes macroscopic. 

\vspace{-2mm}
\subsection{Organizing geometry and decision landscape}

\begin{figure*}[t!]
\centering
\includegraphics[width=0.97\textwidth]{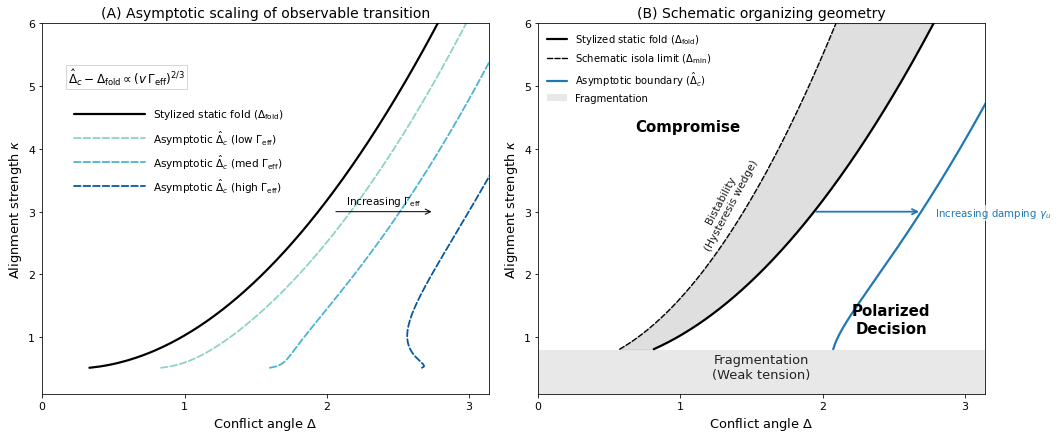}
\vspace{-2mm}
\caption{\textbf{Semi-analytical synthesis of the collective decision landscape and organizing geometry.}
\textbf{(A) Asymptotic scaling of the observable transition.}
The solid black line represents a stylized static fold, $\Delta_{\mathrm{fold}}$, characterizing the structural capacity of the network to sustain compromise. The dashed curves show the asymptotic observable threshold $\hat{\Delta}_c$ predicted by the dynamic bifurcation-delay mechanism derived in SI in Proposition 4, illustrating the delayed onset of polarization. Increasing the effective macroscopic damping $\Gamma_{\mathrm{eff}}$ systematically shifts the observable transition outward according to the scaling law $\hat{\Delta}_c-\Delta_{\mathrm{fold}} \propto (v\Gamma_{\mathrm{eff}})^{2/3}$.
\textbf{(B) Schematic organizing geometry.}
The qualitative arrangement of the stylized static fold $\Delta_{\mathrm{fold}}$, the schematic lower continuation limit $\Delta_{\min}$,  and the fragmentation regime defines the global structure of the decision landscape. Their geometric overlap delineates a hysteresis wedge capable of supporting bistability. During a dynamic sweep, delayed escape from the ghost of the ruptured compromise state shifts the observable transition $\hat{\Delta}_c$ outward under increasing uninformed damping. The figure therefore distinguishes the static structural limits of the network from the finite-time observable boundary that governs when macroscopic polarization becomes visible.}
\label{fig:global_geometry}
\end{figure*}

While Section \ref{3.3} demonstrated topological rupture and dynamic delay along a one-dimensional slice of increasing conflict ($\Delta$), real biological and engineered collectives operate across a multi-dimensional parameter space where internal alignment strength ($\kappa$) also varies. To understand how the local ghost-trapping mechanism scales up to reshape the macroscopic behavior of the group, we must synthesize the global decision landscape. This requires distinguishing the rigid, static topological boundaries of the network from the flexible, observable boundaries dictated by finite-time dynamics and dissipation.
A useful schematic description of the decision landscape in the $(\Delta,\kappa)$ plane involves two distinct structural boundaries, as illustrated in Fig.~\ref{fig:global_geometry}B. The first is the static fold curve, $\Delta_{\mathrm{fold}}(\kappa)$, which marks the structural limit of the compromise branch in the modular regime. The second is a lower continuation limit, denoted $\Delta_{\min}(\kappa)$, associated in the numerics with the disappearance of polarized decision branches.

In the modular configurations studied numerically here, the capacity required to maintain a highly polarized state is lower than that required for global compromise, so that $\Delta_{\mathrm{min}}(\kappa)$ lies to the left of $\Delta_{\mathrm{fold}}(\kappa)$. This topological overlap creates a robust hysteresis wedge of bistability. At sufficiently low alignment strength, the network may also enter a low-global-order regime in which module-level coherence persists while global synchronization is lost. In Fig.~\ref{fig:global_geometry}B this is represented schematically as a fragmentation regime, consistent with the macroscopic definitions introduced in Section~\ref{2.3}, but we do not attempt to derive its boundary analytically here.
However, in continuously varying environments, these static structural boundaries are practically invisible. As established by the dynamic saddle-node reduction (SI Sec. D), finite-time observations record the macroscopic jump at an effective observable threshold $\hat{\Delta}_c(\kappa)$ strictly greater than the static fold. Figure \ref{fig:global_geometry}A displays the asymptotic scaling of this dynamic transition. Rather than a sharp, fixed structural boundary, the delayed onset of polarization follows the theoretical scaling law $\hat{\Delta}_c - \Delta_{\mathrm{fold}} \propto (v \Gamma_{\mathrm{eff}})^{2/3}$, where $v$ is the rate of increasing conflict and $\Gamma_{\mathrm{eff}}$ is the effective damping of the critical network mode.

Because passive participants ($\alpha_i \approx 0$) contribute direction-free dissipation ($\gamma_i > 0$) directly to the longitudinal deformation modes of the network, they act as viscous dampers exactly where the restoring and driving forces nearly balance. Consequently, increasing the effective damping $\Gamma_{\mathrm{eff}}$, for instance, by increasing the fraction of uninformed agents or their local damping $\gamma_u$ dynamically inflates the observable compromise region, shifting $\hat{\Delta}_c(\kappa)$ outward across the parameter space (Fig.~\ref{fig:global_geometry}A and B).

Thus, uninformed individuals provide a robust stabilizing mechanism in this class of structured networks. While uninformed agents do not alter the underlying interaction topology or the static structural limits of the network, they systematically enlarge the observable compromise regime by acting as macroscopic dampers against collective rupture.

\section{Robotic swarms}

To illustrate how the presence of direction-free agents manifests in a noisy
physical system, we analyze data from distributed decision-making experiments
using a swarm of $N=100$ Kilobots with two identical competing sites
\cite{valentini2016collective}. At any given time, a robot occupies one of three discrete internal states: signalling preference for site A, signalling preference for site B, or not signalling any preference. These non-signalling robots serve as a macroscopic experimental analogue of the uninformed agents in our theoretical model. While our framework is formulated in terms of continuous phases on a structured network and the robots operate through discrete states and local broadcasts, the non-signalling robots play an equivalent functional role, they participate in the physical and communication network of the swarm, but inject no directional information.

The data are taken from the University of Bristol dataset
\emph{Robust Distributed Decision-Making in Robot Swarms}
\cite{crosscombeDataset}, associated with the
Kilobot experiments reported in
\cite{crosscombe2017robust}. We analyze $10$
independent experimental runs from the three-valued Kilobot experiments,
recording the time evolution of the number of robots signalling each site
over approximately $2{,}500$ iterations. Rather than reconstructing
individual trajectories or interaction networks, we focus on a
coarse-grained observable that is directly accessible experimentally and
closely mirrors the collective decision observables used in our
theoretical analysis. Let $n_A(t)$ and $n_B(t)$ denote the number of robots signalling preference for sites A and B at time $t$, respectively. We define the macroscopic decision bias as
\begin{equation}
B(t)=\frac{n_A(t)-n_B(t)}{n_A(t)+n_B(t)}.
\end{equation}
Values $|B(t)|\approx 0$ correspond to a lack of consensus, whereas values approaching unity indicate a polarized collective decision.

\begin{figure}[h!]
\centering
\includegraphics[width=1\linewidth]{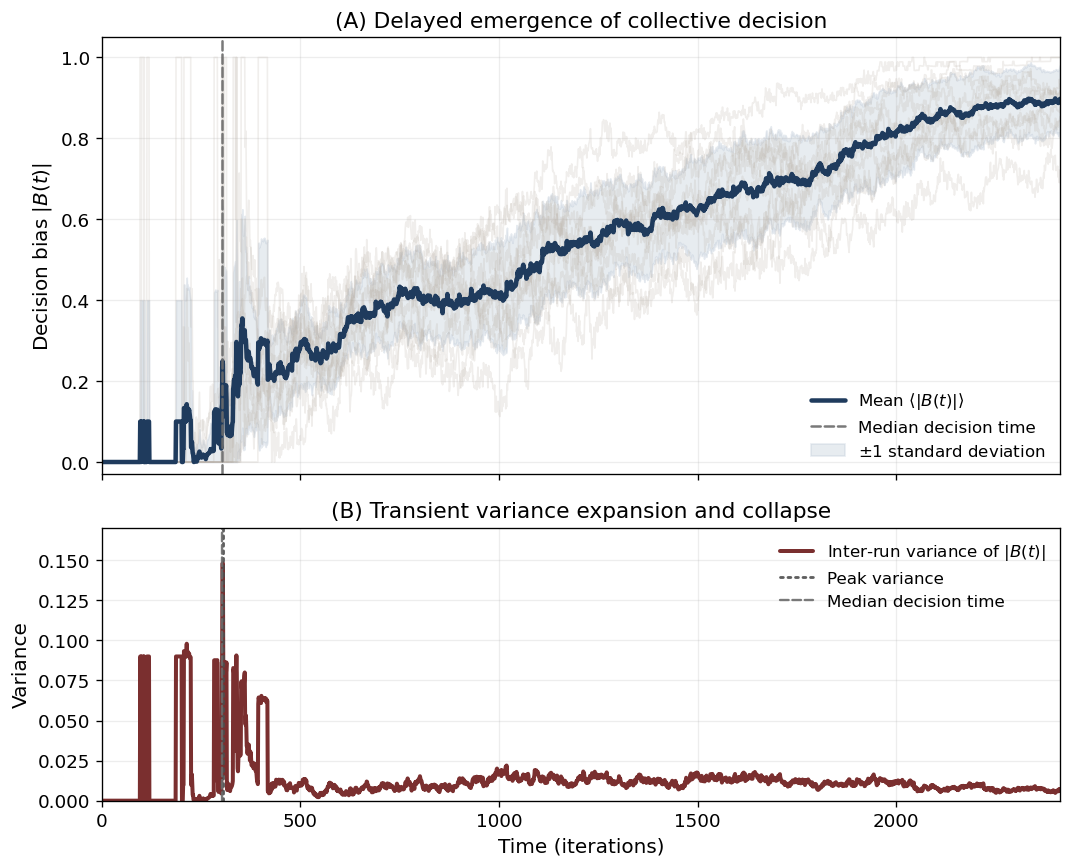}
\caption{\textbf{Delayed decision onset and variance collapse in a robotic swarm.}
Time evolution of the collective decision bias $|B(t)|$ in a swarm of $N=100$ Kilobots containing non-signalling agents. (A) Mean decision bias across $10$ runs (solid line) with the corresponding $\pm 1$ standard deviation band (shaded region). The dashed line marks the median run-level decision time, defined by the first passage to $|B(t)|\ge 0.5$. (B) Inter-run variance of $|B(t)|$, which peaks near $t\approx 305$ and then collapses as the runs converge toward a polarized collective state. This transient expansion and collapse of variance is qualitatively consistent with slowing near a collective decision transition and with delayed escape through a bottleneck-like region.}
\label{fig:kilobots_bias}
\end{figure}

Figure~\ref{fig:kilobots_bias} shows the time evolution of the macroscopic bias $|B(t)|$ averaged over the $10$ experimental runs (panel A), together with the corresponding inter-run variance (panel B). The dynamics exhibit two features that are strongly qualitatively consistent with the delayed decision mechanism identified in Section~3.3. First, the decision bias does not emerge as an abrupt early jump; rather, it develops through a broad transient over many hundreds of iterations before converging toward a strongly polarized state. Second, the inter-run variance undergoes a marked transient expansion, reaches its maximum at approximately $t=305$, and then collapses as the swarm converges toward a coherent collective outcome.

A particularly informative feature is that the peak variance nearly coincides with the typical decision time across runs. Defining a run-level decision time as the first passage of the macroscopic bias through the threshold $|B(t)|\ge 0.5$, we find a median first-passage time of approximately $t\approx300$ (see Fig.~\ref{fig:kilobots_firstpassage_si}). Thus, the ensemble reaches maximal dispersion almost exactly when a typical run first crosses into a macroscopically polarized regime. This close alignment supports the interpretation that the swarm traverses a broad high-susceptibility transient before resolving into a collective decision.
In the language of our theoretical framework, the high-variance stage is qualitatively consistent with delayed traversal of a bottleneck-like region in the decision landscape. During this transient, the effective restoring forces toward any particular option remain weak, leaving the swarm more susceptible to stochastic fluctuations and finite-size effects. Once the system escapes this bottleneck and moves toward a polarized collective state, trajectories bundle more tightly, leading to the observed variance collapse.

These data are qualitatively consistent with the dissipation-delay mechanism proposed here. Alternative dynamical scenarios may produce similar signatures;  a full empirical study, which may disambiguate the mechanisms, is beyond the scope of this work. We thus regard the robotic data as supportive rather than decisive.

\section{Discussion}
\label{sec:discussion}

This work provides a geometric and dynamical explanation for a phenomenon observed across biological and engineered collectives. The stabilizing role of uninformed agents in collective decision-making. The central point is not simply that passive participants slow the dynamics, but that in the structured second-order networks studied here they reshape both the mechanism by which compromise is lost and the conflict level at which polarization becomes macroscopic.

\subsection{Topological rupture of compromise in structured networks}

A first conceptual outcome is that compromise loss in structured networks need not follow the smooth symmetry-breaking scenario familiar from mean-field models \cite{nabet2009dynamics}. In the modular setting studied here, the compromise state is instead a network-constrained collective configuration whose phase tension becomes concentrated across sparse bridges and cut-sets.
 Because pairwise alignment forces are bounded, these bottlenecks impose a finite structural capacity. As conflict increases, the compromise branch can therefore terminate through a saddle-node fold rather than a local pitchfork-type transition.

This distinction matters because it reorganizes the global decision landscape. Once the compromise branch terminates at the fold, the polarized decision states cannot emerge locally from it.  Instead, in the numerical continuation picture studied here, they occupy structurally separated branches, yielding a hysteresis wedge between the fold of compromise and the lower continuation limit of the decision branches. In this sense, the relevant geometry is not one of smooth branch splitting, but one of topological rupture, disconnected equilibria, and finite-capacity transmission of alignment across the network.

\vspace{-2mm}

\subsection{Uninformed agents reshape the observable transition}

A second central result is that uninformed individuals act as dampers of the critical slow mode that governs the approach to rupture. Informed leaders inject directional forcing into the same weak longitudinal modes through which conflict propagates. Uninformed agents, although they contribute no directional bias, inject dissipation directly into these modes. This does not alter the static topological capacity of the network, with the structural fold remaining fixed. What changes is the dynamical time scale on which rupture becomes observable.

Near the fold, loss of the compromise equilibrium gives rise to a saddle-node ghost that can trap trajectories in a long-lived bottleneck.
 Increasing direction-free dissipation prolongs this traversal and therefore delays the finite-time onset of visible polarization. The consequence is a clear separation between the static structural threshold and the effective observable threshold recorded under finite-time observation or continuous environmental ramping. Passive participants thus do not prevent rupture, but they can strongly mask its onset and enlarge the functional regime in which the group appears to maintain compromise.

This perspective also clarifies the role of the global energy-dissipation balance. Because alignment interactions are power-preserving, they can redistribute activity across the network but cannot create it. Irreversible contraction of the decision landscape arises only through dissipation. The stabilizing effect of uninformed agents is therefore not merely combinatorial or connectivity-based, it is dynamical, modal, and energy-constrained.

\vspace{-2mm}
\subsection{Extensions}

Our framework suggests a new dynamical interpretation of prior biological observations on uninformed individuals in collective choice. In particular, the classic experiments of Couzin et al.~\cite{couzin2011uninformed} showed that uninformed fish can promote democratic consensus in groups facing competing directional preferences. Within our geometric picture, such individuals need not merely rebalance influence or average local signals; they may also delay premature commitment by increasing effective dissipation along the slow collective modes that organize the loss of compromise. Although the available fish data do not permit a direct quantitative test of the delay law derived here, they are qualitatively consistent with the broader possibility that direction-free participants stabilize collective decisions not only through network connectivity, but also through dynamical retardation of symmetry breaking.

The robotic swarm experiment provides qualitative consistency with this mechanism. In the Kilobot system, collective decision formation does not occur as an immediate jump, but through a broad transient in which inter-run variance expands, peaks near the median decision time, and then collapses as the swarm converges toward a polarized macroscopic outcome. We do not treat these data as a quantitative validation of the delay law, but rather as a proof-of-concept that the predicted transient signature can arise in a physical collective containing direction-free participants.
Notably, dedicated experiments could further test the mechanism presented here, for instance one might seek to empirically identify the effective damping of slow collective modes directly from trajectory data in animal groups or robot swarms.

Several extensions follow naturally. First, the present analysis focuses on two competing options, but the geometric framework is not restricted to this case and should extend to multiple competing directions. Second, we have treated the interaction graph as static, whereas many biological and engineered collectives evolve on adaptive or time-varying networks. Extending the theory to dynamic network architectures would clarify how topological rupture and dynamical masking interact with state-dependent communication. Third, while the present model includes memory and dissipation through second-order dynamics, it does not explicitly incorporate stochastic forcing at the agent level. Noise is likely to interact nontrivially with the fold geometry and ghost-induced delay, particularly near the observable transition. 
A further natural extension would be to consider the balanced mean-field regime, where the corresponding loss of compromise may be organized by a pitchfork-type critical-mode reduction.

Overall, our results show that uninformed individuals play a constructive dynamical role in collective decision-making. By injecting dissipation without directional bias, they stabilize compromise at low conflict, delay the observable onset of polarization near rupture, and reshape the global decision landscape in structured networks. More generally, the work shows how geometric mechanics and energy-based modeling can reveal organizing principles of collective behavior that remain hidden in purely connectivity-based descriptions.


\vspace{6mm}{\bf Acknowledgments.}
L. C acknowledges financial support from Grant PID2022-137909NB-C21 funded by MCIN/AEI/ 10.13039/501100011033 and  iRoboCity2030-CM, TEC-2024/TEC-62.  The research of LPS is partially supported by  NSF FRG Award DMS- 2152107 and NSF CAREER Award DMS 1749013, and also supported in part by grants from the NSF (DMS-2235451) and Simons
Foundation (MP-TMPS-00005320) to the NSF-Simons National Institute for Theory and
Mathematics in Biology (NITMB). JU is supported by NSF Award PHY-2209998.



\vspace{5mm}{\bf Code and data availability.}
Codes, parameter files, and curated data used to generate the figures and numerical results in this paper
are available on request.

\vspace{3mm}

\bibliography{mybib}

\begin{thebibliography}{31}%
\makeatletter
\providecommand \@ifxundefined [1]{%
 \@ifx{#1\undefined}
}%
\providecommand \@ifnum [1]{%
 \ifnum #1\expandafter \@firstoftwo
 \else \expandafter \@secondoftwo
 \fi
}%
\providecommand \@ifx [1]{%
 \ifx #1\expandafter \@firstoftwo
 \else \expandafter \@secondoftwo
 \fi
}%
\providecommand \natexlab [1]{#1}%
\providecommand \enquote  [1]{``#1''}%
\providecommand \bibnamefont  [1]{#1}%
\providecommand \bibfnamefont [1]{#1}%
\providecommand \citenamefont [1]{#1}%
\providecommand \href@noop [0]{\@secondoftwo}%
\providecommand \href [0]{\begingroup \@sanitize@url \@href}%
\providecommand \@href[1]{\@@startlink{#1}\@@href}%
\providecommand \@@href[1]{\endgroup#1\@@endlink}%
\providecommand \@sanitize@url [0]{\catcode `\\12\catcode `\$12\catcode
  `\&12\catcode `\#12\catcode `\^12\catcode `\_12\catcode `\%12\relax}%
\providecommand \@@startlink[1]{}%
\providecommand \@@endlink[0]{}%
\providecommand \url  [0]{\begingroup\@sanitize@url \@url }%
\providecommand \@url [1]{\endgroup\@href {#1}{\urlprefix }}%
\providecommand \urlprefix  [0]{URL }%
\providecommand \Eprint [0]{\href }%
\providecommand \doibase [0]{https://doi.org/}%
\providecommand \selectlanguage [0]{\@gobble}%
\providecommand \bibinfo  [0]{\@secondoftwo}%
\providecommand \bibfield  [0]{\@secondoftwo}%
\providecommand \translation [1]{[#1]}%
\providecommand \BibitemOpen [0]{}%
\providecommand \bibitemStop [0]{}%
\providecommand \bibitemNoStop [0]{.\EOS\space}%
\providecommand \EOS [0]{\spacefactor3000\relax}%
\providecommand \BibitemShut  [1]{\csname bibitem#1\endcsname}%
\let\auto@bib@innerbib\@empty
\bibitem [{\citenamefont {Couzin}\ \emph {et~al.}(2005)\citenamefont {Couzin},
  \citenamefont {Krause}, \citenamefont {Franks},\ and\ \citenamefont
  {Levin}}]{couzin2005effective}%
  \BibitemOpen
  \bibfield  {author} {\bibinfo {author} {\bibfnamefont {I.~D.}\ \bibnamefont
  {Couzin}}, \bibinfo {author} {\bibfnamefont {J.}~\bibnamefont {Krause}},
  \bibinfo {author} {\bibfnamefont {N.~R.}\ \bibnamefont {Franks}},\ and\
  \bibinfo {author} {\bibfnamefont {S.~A.}\ \bibnamefont {Levin}},\ }\bibfield
  {title} {\bibinfo {title} {Effective leadership and decision-making in animal
  groups on the move},\ }\href@noop {} {\bibfield  {journal} {\bibinfo
  {journal} {Nature}\ }\textbf {\bibinfo {volume} {433}},\ \bibinfo {pages}
  {513} (\bibinfo {year} {2005})}\BibitemShut {NoStop}%
\bibitem [{\citenamefont {Leonard}\ \emph {et~al.}(2012)\citenamefont
  {Leonard}, \citenamefont {Shen}, \citenamefont {Nabet}, \citenamefont
  {Scardovi}, \citenamefont {Couzin},\ and\ \citenamefont
  {Levin}}]{leonard2012decision}%
  \BibitemOpen
  \bibfield  {author} {\bibinfo {author} {\bibfnamefont {N.~E.}\ \bibnamefont
  {Leonard}}, \bibinfo {author} {\bibfnamefont {T.}~\bibnamefont {Shen}},
  \bibinfo {author} {\bibfnamefont {B.}~\bibnamefont {Nabet}}, \bibinfo
  {author} {\bibfnamefont {L.}~\bibnamefont {Scardovi}}, \bibinfo {author}
  {\bibfnamefont {I.~D.}\ \bibnamefont {Couzin}},\ and\ \bibinfo {author}
  {\bibfnamefont {S.~A.}\ \bibnamefont {Levin}},\ }\bibfield  {title} {\bibinfo
  {title} {Decision versus compromise for animal groups in motion},\
  }\href@noop {} {\bibfield  {journal} {\bibinfo  {journal} {Proceedings of the
  National Academy of Sciences}\ }\textbf {\bibinfo {volume} {109}},\ \bibinfo
  {pages} {227} (\bibinfo {year} {2012})}\BibitemShut {NoStop}%
\bibitem [{\citenamefont {Herbert-Read}\ \emph {et~al.}(2011)\citenamefont
  {Herbert-Read}, \citenamefont {Perna}, \citenamefont {Mann}, \citenamefont
  {Schaerf}, \citenamefont {Sumpter},\ and\ \citenamefont
  {Ward}}]{herbert2011inferring}%
  \BibitemOpen
  \bibfield  {author} {\bibinfo {author} {\bibfnamefont {J.~E.}\ \bibnamefont
  {Herbert-Read}}, \bibinfo {author} {\bibfnamefont {A.}~\bibnamefont {Perna}},
  \bibinfo {author} {\bibfnamefont {R.~P.}\ \bibnamefont {Mann}}, \bibinfo
  {author} {\bibfnamefont {T.~M.}\ \bibnamefont {Schaerf}}, \bibinfo {author}
  {\bibfnamefont {D.~J.}\ \bibnamefont {Sumpter}},\ and\ \bibinfo {author}
  {\bibfnamefont {A.~J.}\ \bibnamefont {Ward}},\ }\bibfield  {title} {\bibinfo
  {title} {Inferring the rules of interaction of shoaling fish},\ }\href@noop
  {} {\bibfield  {journal} {\bibinfo  {journal} {Proceedings of the National
  Academy of Sciences}\ }\textbf {\bibinfo {volume} {108}},\ \bibinfo {pages}
  {18726} (\bibinfo {year} {2011})}\BibitemShut {NoStop}%
\bibitem [{\citenamefont {Krause}\ and\ \citenamefont
  {Ruxton}(2002)}]{krause2002living}%
  \BibitemOpen
  \bibfield  {author} {\bibinfo {author} {\bibfnamefont {J.}~\bibnamefont
  {Krause}}\ and\ \bibinfo {author} {\bibfnamefont {G.~D.}\ \bibnamefont
  {Ruxton}},\ }\href@noop {} {\emph {\bibinfo {title} {Living in groups}}}\
  (\bibinfo  {publisher} {Oxford University Press},\ \bibinfo {year}
  {2002})\BibitemShut {NoStop}%
\bibitem [{\citenamefont {Conradt}\ and\ \citenamefont
  {Roper}(2005)}]{conradt2005consensus}%
  \BibitemOpen
  \bibfield  {author} {\bibinfo {author} {\bibfnamefont {L.}~\bibnamefont
  {Conradt}}\ and\ \bibinfo {author} {\bibfnamefont {T.~J.}\ \bibnamefont
  {Roper}},\ }\bibfield  {title} {\bibinfo {title} {Consensus decision making
  in animals},\ }\href@noop {} {\bibfield  {journal} {\bibinfo  {journal}
  {Trends in ecology \& evolution}\ }\textbf {\bibinfo {volume} {20}},\
  \bibinfo {pages} {449} (\bibinfo {year} {2005})}\BibitemShut {NoStop}%
\bibitem [{\citenamefont {Sumpter}\ \emph {et~al.}(2008)\citenamefont
  {Sumpter}, \citenamefont {Krause}, \citenamefont {James}, \citenamefont
  {Couzin},\ and\ \citenamefont {Ward}}]{sumpter2008consensus}%
  \BibitemOpen
  \bibfield  {author} {\bibinfo {author} {\bibfnamefont {D.~J.}\ \bibnamefont
  {Sumpter}}, \bibinfo {author} {\bibfnamefont {J.}~\bibnamefont {Krause}},
  \bibinfo {author} {\bibfnamefont {R.}~\bibnamefont {James}}, \bibinfo
  {author} {\bibfnamefont {I.~D.}\ \bibnamefont {Couzin}},\ and\ \bibinfo
  {author} {\bibfnamefont {A.~J.}\ \bibnamefont {Ward}},\ }\bibfield  {title}
  {\bibinfo {title} {Consensus decision making by fish},\ }\href@noop {}
  {\bibfield  {journal} {\bibinfo  {journal} {Current Biology}\ }\textbf
  {\bibinfo {volume} {18}},\ \bibinfo {pages} {1773} (\bibinfo {year}
  {2008})}\BibitemShut {NoStop}%
\bibitem [{\citenamefont {Couzin}\ \emph {et~al.}(2011)\citenamefont {Couzin},
  \citenamefont {Ioannou}, \citenamefont {Demirel}, \citenamefont {Gross},
  \citenamefont {Torney}, \citenamefont {Hartnett}, \citenamefont {Conradt},
  \citenamefont {Levin},\ and\ \citenamefont {Leonard}}]{couzin2011uninformed}%
  \BibitemOpen
  \bibfield  {author} {\bibinfo {author} {\bibfnamefont {I.~D.}\ \bibnamefont
  {Couzin}}, \bibinfo {author} {\bibfnamefont {C.~C.}\ \bibnamefont {Ioannou}},
  \bibinfo {author} {\bibfnamefont {G.}~\bibnamefont {Demirel}}, \bibinfo
  {author} {\bibfnamefont {T.}~\bibnamefont {Gross}}, \bibinfo {author}
  {\bibfnamefont {C.~J.}\ \bibnamefont {Torney}}, \bibinfo {author}
  {\bibfnamefont {A.}~\bibnamefont {Hartnett}}, \bibinfo {author}
  {\bibfnamefont {L.}~\bibnamefont {Conradt}}, \bibinfo {author} {\bibfnamefont
  {S.~A.}\ \bibnamefont {Levin}},\ and\ \bibinfo {author} {\bibfnamefont
  {N.~E.}\ \bibnamefont {Leonard}},\ }\bibfield  {title} {\bibinfo {title}
  {Uninformed individuals promote democratic consensus in animal groups},\
  }\href@noop {} {\bibfield  {journal} {\bibinfo  {journal} {science}\ }\textbf
  {\bibinfo {volume} {334}},\ \bibinfo {pages} {1578} (\bibinfo {year}
  {2011})}\BibitemShut {NoStop}%
\bibitem [{\citenamefont {Vicsek}\ \emph {et~al.}(1995)\citenamefont {Vicsek},
  \citenamefont {Czir{\'o}k}, \citenamefont {Ben-Jacob}, \citenamefont
  {Cohen},\ and\ \citenamefont {Shochet}}]{vicsek1995novel}%
  \BibitemOpen
  \bibfield  {author} {\bibinfo {author} {\bibfnamefont {T.}~\bibnamefont
  {Vicsek}}, \bibinfo {author} {\bibfnamefont {A.}~\bibnamefont {Czir{\'o}k}},
  \bibinfo {author} {\bibfnamefont {E.}~\bibnamefont {Ben-Jacob}}, \bibinfo
  {author} {\bibfnamefont {I.}~\bibnamefont {Cohen}},\ and\ \bibinfo {author}
  {\bibfnamefont {O.}~\bibnamefont {Shochet}},\ }\bibfield  {title} {\bibinfo
  {title} {Novel type of phase transition in a system of self-driven
  particles},\ }\href@noop {} {\bibfield  {journal} {\bibinfo  {journal}
  {Physical review letters}\ }\textbf {\bibinfo {volume} {75}},\ \bibinfo
  {pages} {1226} (\bibinfo {year} {1995})}\BibitemShut {NoStop}%
\bibitem [{\citenamefont {Yates}\ \emph {et~al.}(2009)\citenamefont {Yates},
  \citenamefont {Erban}, \citenamefont {Escudero}, \citenamefont {Couzin},
  \citenamefont {Buhl}, \citenamefont {Kevrekidis}, \citenamefont {Maini},\
  and\ \citenamefont {Sumpter}}]{yates2009inherent}%
  \BibitemOpen
  \bibfield  {author} {\bibinfo {author} {\bibfnamefont {C.~A.}\ \bibnamefont
  {Yates}}, \bibinfo {author} {\bibfnamefont {R.}~\bibnamefont {Erban}},
  \bibinfo {author} {\bibfnamefont {C.}~\bibnamefont {Escudero}}, \bibinfo
  {author} {\bibfnamefont {I.~D.}\ \bibnamefont {Couzin}}, \bibinfo {author}
  {\bibfnamefont {C.}~\bibnamefont {Buhl}}, \bibinfo {author} {\bibfnamefont
  {I.~G.}\ \bibnamefont {Kevrekidis}}, \bibinfo {author} {\bibfnamefont
  {P.~K.}\ \bibnamefont {Maini}},\ and\ \bibinfo {author} {\bibfnamefont
  {D.~J.}\ \bibnamefont {Sumpter}},\ }\bibfield  {title} {\bibinfo {title}
  {Inherent noise can facilitate coherence in collective swarm motion},\
  }\href@noop {} {\bibfield  {journal} {\bibinfo  {journal} {Proceedings of the
  National Academy of Sciences}\ }\textbf {\bibinfo {volume} {106}},\ \bibinfo
  {pages} {5464} (\bibinfo {year} {2009})}\BibitemShut {NoStop}%
\bibitem [{\citenamefont {Arganda}\ \emph {et~al.}(2012)\citenamefont
  {Arganda}, \citenamefont {P{\'e}rez-Escudero},\ and\ \citenamefont
  {de~Polavieja}}]{arganda2012common}%
  \BibitemOpen
  \bibfield  {author} {\bibinfo {author} {\bibfnamefont {S.}~\bibnamefont
  {Arganda}}, \bibinfo {author} {\bibfnamefont {A.}~\bibnamefont
  {P{\'e}rez-Escudero}},\ and\ \bibinfo {author} {\bibfnamefont {G.~G.}\
  \bibnamefont {de~Polavieja}},\ }\bibfield  {title} {\bibinfo {title} {A
  common rule for decision making in animal collectives across species},\
  }\href@noop {} {\bibfield  {journal} {\bibinfo  {journal} {Proceedings of the
  National Academy of Sciences}\ }\textbf {\bibinfo {volume} {109}},\ \bibinfo
  {pages} {20508} (\bibinfo {year} {2012})}\BibitemShut {NoStop}%
\bibitem [{\citenamefont {Marshall}\ \emph {et~al.}(2009)\citenamefont
  {Marshall}, \citenamefont {Bogacz}, \citenamefont {Dornhaus}, \citenamefont
  {Planqu{\'e}}, \citenamefont {Kovacs},\ and\ \citenamefont
  {Franks}}]{marshall2009optimal}%
  \BibitemOpen
  \bibfield  {author} {\bibinfo {author} {\bibfnamefont {J.~A.}\ \bibnamefont
  {Marshall}}, \bibinfo {author} {\bibfnamefont {R.}~\bibnamefont {Bogacz}},
  \bibinfo {author} {\bibfnamefont {A.}~\bibnamefont {Dornhaus}}, \bibinfo
  {author} {\bibfnamefont {R.}~\bibnamefont {Planqu{\'e}}}, \bibinfo {author}
  {\bibfnamefont {T.}~\bibnamefont {Kovacs}},\ and\ \bibinfo {author}
  {\bibfnamefont {N.~R.}\ \bibnamefont {Franks}},\ }\bibfield  {title}
  {\bibinfo {title} {On optimal decision-making in brains and social insect
  colonies},\ }\href@noop {} {\bibfield  {journal} {\bibinfo  {journal}
  {Journal of the Royal Society Interface}\ }\textbf {\bibinfo {volume} {6}},\
  \bibinfo {pages} {1065} (\bibinfo {year} {2009})}\BibitemShut {NoStop}%
\bibitem [{\citenamefont {Deneubourg}\ \emph {et~al.}(1990)\citenamefont
  {Deneubourg}, \citenamefont {Aron}, \citenamefont {Goss},\ and\ \citenamefont
  {Pasteels}}]{deneubourg1990self}%
  \BibitemOpen
  \bibfield  {author} {\bibinfo {author} {\bibfnamefont {J.-L.}\ \bibnamefont
  {Deneubourg}}, \bibinfo {author} {\bibfnamefont {S.}~\bibnamefont {Aron}},
  \bibinfo {author} {\bibfnamefont {S.}~\bibnamefont {Goss}},\ and\ \bibinfo
  {author} {\bibfnamefont {J.~M.}\ \bibnamefont {Pasteels}},\ }\bibfield
  {title} {\bibinfo {title} {The self-organizing exploratory pattern of the
  argentine ant},\ }\href@noop {} {\bibfield  {journal} {\bibinfo  {journal}
  {Journal of insect behavior}\ }\textbf {\bibinfo {volume} {3}},\ \bibinfo
  {pages} {159} (\bibinfo {year} {1990})}\BibitemShut {NoStop}%
\bibitem [{\citenamefont {Gold}\ and\ \citenamefont
  {Shadlen}(2007)}]{gold2007neural}%
  \BibitemOpen
  \bibfield  {author} {\bibinfo {author} {\bibfnamefont {J.~I.}\ \bibnamefont
  {Gold}}\ and\ \bibinfo {author} {\bibfnamefont {M.~N.}\ \bibnamefont
  {Shadlen}},\ }\bibfield  {title} {\bibinfo {title} {The neural basis of
  decision making},\ }\href@noop {} {\bibfield  {journal} {\bibinfo  {journal}
  {Annu. Rev. Neurosci.}\ }\textbf {\bibinfo {volume} {30}},\ \bibinfo {pages}
  {535} (\bibinfo {year} {2007})}\BibitemShut {NoStop}%
\bibitem [{\citenamefont {Ratcliff}\ and\ \citenamefont
  {McKoon}(2008)}]{ratcliff2008diffusion}%
  \BibitemOpen
  \bibfield  {author} {\bibinfo {author} {\bibfnamefont {R.}~\bibnamefont
  {Ratcliff}}\ and\ \bibinfo {author} {\bibfnamefont {G.}~\bibnamefont
  {McKoon}},\ }\bibfield  {title} {\bibinfo {title} {The diffusion decision
  model: theory and data for two-choice decision tasks},\ }\href@noop {}
  {\bibfield  {journal} {\bibinfo  {journal} {Neural computation}\ }\textbf
  {\bibinfo {volume} {20}},\ \bibinfo {pages} {873} (\bibinfo {year}
  {2008})}\BibitemShut {NoStop}%
\bibitem [{\citenamefont {Kuramoto}(1975)}]{kuramoto1975international}%
  \BibitemOpen
  \bibfield  {author} {\bibinfo {author} {\bibfnamefont {Y.}~\bibnamefont
  {Kuramoto}},\ }\bibfield  {title} {\bibinfo {title} {International symposium
  on mathematical problems in theoretical physics},\ }\href@noop {} {\bibfield
  {journal} {\bibinfo  {journal} {Lecture notes in Physics}\ }\textbf {\bibinfo
  {volume} {30}},\ \bibinfo {pages} {420} (\bibinfo {year} {1975})}\BibitemShut
  {NoStop}%
\bibitem [{\citenamefont {Grason}(2016)}]{grason2016perspective}%
  \BibitemOpen
  \bibfield  {author} {\bibinfo {author} {\bibfnamefont {G.~M.}\ \bibnamefont
  {Grason}},\ }\bibfield  {title} {\bibinfo {title} {Perspective: Geometrically
  frustrated assemblies},\ }\href@noop {} {\bibfield  {journal} {\bibinfo
  {journal} {The Journal of Chemical Physics}\ }\textbf {\bibinfo {volume}
  {145}} (\bibinfo {year} {2016})}\BibitemShut {NoStop}%
\bibitem [{\citenamefont {Giampaolo}\ \emph {et~al.}(2011)\citenamefont
  {Giampaolo}, \citenamefont {Gualdi}, \citenamefont {Monras},\ and\
  \citenamefont {Illuminati}}]{Giampaolo:2011fqf}%
  \BibitemOpen
  \bibfield  {author} {\bibinfo {author} {\bibfnamefont {S.~M.}\ \bibnamefont
  {Giampaolo}}, \bibinfo {author} {\bibfnamefont {G.}~\bibnamefont {Gualdi}},
  \bibinfo {author} {\bibfnamefont {A.}~\bibnamefont {Monras}},\ and\ \bibinfo
  {author} {\bibfnamefont {F.}~\bibnamefont {Illuminati}},\ }\bibfield  {title}
  {\bibinfo {title} {{Characterizing and Quantifying Frustration in Quantum
  Many-Body Systems}},\ }\href {https://doi.org/10.1103/PhysRevLett.107.260602}
  {\bibfield  {journal} {\bibinfo  {journal} {Phys. Rev. Lett.}\ }\textbf
  {\bibinfo {volume} {107}},\ \bibinfo {pages} {260602} (\bibinfo {year}
  {2011})},\ \Eprint {https://arxiv.org/abs/1103.0022} {arXiv:1103.0022
  [cond-mat.other]} \BibitemShut {NoStop}%
\bibitem [{\citenamefont {Nabet}\ \emph {et~al.}(2009)\citenamefont {Nabet},
  \citenamefont {Leonard}, \citenamefont {Couzin},\ and\ \citenamefont
  {Levin}}]{nabet2009dynamics}%
  \BibitemOpen
  \bibfield  {author} {\bibinfo {author} {\bibfnamefont {B.}~\bibnamefont
  {Nabet}}, \bibinfo {author} {\bibfnamefont {N.~E.}\ \bibnamefont {Leonard}},
  \bibinfo {author} {\bibfnamefont {I.~D.}\ \bibnamefont {Couzin}},\ and\
  \bibinfo {author} {\bibfnamefont {S.~A.}\ \bibnamefont {Levin}},\ }\bibfield
  {title} {\bibinfo {title} {Dynamics of decision making in animal group
  motion},\ }\href@noop {} {\bibfield  {journal} {\bibinfo  {journal} {Journal
  of nonlinear science}\ }\textbf {\bibinfo {volume} {19}},\ \bibinfo {pages}
  {399} (\bibinfo {year} {2009})}\BibitemShut {NoStop}%
\bibitem [{\citenamefont {Valentini}\ \emph {et~al.}(2016)\citenamefont
  {Valentini}, \citenamefont {Ferrante}, \citenamefont {Hamann},\ and\
  \citenamefont {Dorigo}}]{valentini2016collective}%
  \BibitemOpen
  \bibfield  {author} {\bibinfo {author} {\bibfnamefont {G.}~\bibnamefont
  {Valentini}}, \bibinfo {author} {\bibfnamefont {E.}~\bibnamefont {Ferrante}},
  \bibinfo {author} {\bibfnamefont {H.}~\bibnamefont {Hamann}},\ and\ \bibinfo
  {author} {\bibfnamefont {M.}~\bibnamefont {Dorigo}},\ }\bibfield  {title}
  {\bibinfo {title} {Collective decision with 100 kilobots: Speed versus
  accuracy in binary discrimination problems},\ }\href@noop {} {\bibfield
  {journal} {\bibinfo  {journal} {Autonomous agents and multi-agent systems}\
  }\textbf {\bibinfo {volume} {30}},\ \bibinfo {pages} {553} (\bibinfo {year}
  {2016})}\BibitemShut {NoStop}%
\bibitem [{\citenamefont {Lawry}\ and\ \citenamefont
  {Crosscombe}(2017)}]{crosscombeDataset}%
  \BibitemOpen
  \bibfield  {author} {\bibinfo {author} {\bibfnamefont {J.}~\bibnamefont
  {Lawry}}\ and\ \bibinfo {author} {\bibfnamefont {M.}~\bibnamefont
  {Crosscombe}},\ }\bibfield  {title} {\bibinfo {title} {Robust distributed
  decision-making in robot swarms},\ }\href
  {https://doi.org/10.5523/bris.2dh47v1ak21vi22o5pn5ki1tpy}
  {10.5523/bris.2dh47v1ak21vi22o5pn5ki1tpy} (\bibinfo {year}
  {2017})\BibitemShut {NoStop}%
\bibitem [{\citenamefont {Crosscombe}\ \emph {et~al.}(2017)\citenamefont
  {Crosscombe}, \citenamefont {Lawry}, \citenamefont {Hauert},\ and\
  \citenamefont {Homer}}]{crosscombe2017robust}%
  \BibitemOpen
  \bibfield  {author} {\bibinfo {author} {\bibfnamefont {M.}~\bibnamefont
  {Crosscombe}}, \bibinfo {author} {\bibfnamefont {J.}~\bibnamefont {Lawry}},
  \bibinfo {author} {\bibfnamefont {S.}~\bibnamefont {Hauert}},\ and\ \bibinfo
  {author} {\bibfnamefont {M.}~\bibnamefont {Homer}},\ }\bibfield  {title}
  {\bibinfo {title} {Robust distributed decision-making in robot swarms:
  Exploiting a third truth state},\ }in\ \href@noop {} {\emph {\bibinfo
  {booktitle} {2017 IEEE/RSJ international conference on intelligent robots and
  systems (IROS)}}}\ (\bibinfo {organization} {IEEE},\ \bibinfo {year} {2017})\
  pp.\ \bibinfo {pages} {4326--4332}\BibitemShut {NoStop}%
\bibitem [{\citenamefont {Olfati-Saber}\ \emph {et~al.}(2007)\citenamefont
  {Olfati-Saber}, \citenamefont {Fax},\ and\ \citenamefont
  {Murray}}]{olfati2007consensus}%
  \BibitemOpen
  \bibfield  {author} {\bibinfo {author} {\bibfnamefont {R.}~\bibnamefont
  {Olfati-Saber}}, \bibinfo {author} {\bibfnamefont {J.~A.}\ \bibnamefont
  {Fax}},\ and\ \bibinfo {author} {\bibfnamefont {R.~M.}\ \bibnamefont
  {Murray}},\ }\bibfield  {title} {\bibinfo {title} {Consensus and cooperation
  in networked multi-agent systems},\ }\href@noop {} {\bibfield  {journal}
  {\bibinfo  {journal} {Proceedings of the IEEE}\ }\textbf {\bibinfo {volume}
  {95}},\ \bibinfo {pages} {215} (\bibinfo {year} {2007})}\BibitemShut
  {NoStop}%
\bibitem [{\citenamefont {Cavagna}\ and\ \citenamefont
  {Giardina}(2014)}]{Cavagna2014bird}%
  \BibitemOpen
  \bibfield  {author} {\bibinfo {author} {\bibfnamefont {A.}~\bibnamefont
  {Cavagna}}\ and\ \bibinfo {author} {\bibfnamefont {I.}~\bibnamefont
  {Giardina}},\ }\bibfield  {title} {\bibinfo {title} {Bird flocks as condensed
  matter},\ }\href@noop {} {\bibfield  {journal} {\bibinfo  {journal} {Annu.
  Rev. Condens. Matter Phys.}\ }\textbf {\bibinfo {volume} {5}},\ \bibinfo
  {pages} {183} (\bibinfo {year} {2014})}\BibitemShut {NoStop}%
\bibitem [{\citenamefont {Attanasi}\ \emph {et~al.}(2014)\citenamefont
  {Attanasi}, \citenamefont {Cavagna}, \citenamefont {Del~Castello},
  \citenamefont {Giardina}, \citenamefont {Melillo}, \citenamefont {Parisi},
  \citenamefont {Pohl}, \citenamefont {Rossaro}, \citenamefont {Shen},
  \citenamefont {Silvestri} \emph {et~al.}}]{attanasi2014collective}%
  \BibitemOpen
  \bibfield  {author} {\bibinfo {author} {\bibfnamefont {A.}~\bibnamefont
  {Attanasi}}, \bibinfo {author} {\bibfnamefont {A.}~\bibnamefont {Cavagna}},
  \bibinfo {author} {\bibfnamefont {L.}~\bibnamefont {Del~Castello}}, \bibinfo
  {author} {\bibfnamefont {I.}~\bibnamefont {Giardina}}, \bibinfo {author}
  {\bibfnamefont {S.}~\bibnamefont {Melillo}}, \bibinfo {author} {\bibfnamefont
  {L.}~\bibnamefont {Parisi}}, \bibinfo {author} {\bibfnamefont
  {O.}~\bibnamefont {Pohl}}, \bibinfo {author} {\bibfnamefont {B.}~\bibnamefont
  {Rossaro}}, \bibinfo {author} {\bibfnamefont {E.}~\bibnamefont {Shen}},
  \bibinfo {author} {\bibfnamefont {E.}~\bibnamefont {Silvestri}}, \emph
  {et~al.},\ }\bibfield  {title} {\bibinfo {title} {Collective behaviour
  without collective order in wild swarms of midges},\ }\href@noop {}
  {\bibfield  {journal} {\bibinfo  {journal} {PLoS computational biology}\
  }\textbf {\bibinfo {volume} {10}},\ \bibinfo {pages} {e1003697} (\bibinfo
  {year} {2014})}\BibitemShut {NoStop}%
\bibitem [{\citenamefont {Katz}\ \emph {et~al.}(2011)\citenamefont {Katz},
  \citenamefont {Tunstr{\o}m}, \citenamefont {Ioannou}, \citenamefont {Huepe},\
  and\ \citenamefont {Couzin}}]{katz2011inferring}%
  \BibitemOpen
  \bibfield  {author} {\bibinfo {author} {\bibfnamefont {Y.}~\bibnamefont
  {Katz}}, \bibinfo {author} {\bibfnamefont {K.}~\bibnamefont {Tunstr{\o}m}},
  \bibinfo {author} {\bibfnamefont {C.~C.}\ \bibnamefont {Ioannou}}, \bibinfo
  {author} {\bibfnamefont {C.}~\bibnamefont {Huepe}},\ and\ \bibinfo {author}
  {\bibfnamefont {I.~D.}\ \bibnamefont {Couzin}},\ }\bibfield  {title}
  {\bibinfo {title} {Inferring the structure and dynamics of interactions in
  schooling fish},\ }\href@noop {} {\bibfield  {journal} {\bibinfo  {journal}
  {Proceedings of the National Academy of Sciences}\ }\textbf {\bibinfo
  {volume} {108}},\ \bibinfo {pages} {18720} (\bibinfo {year}
  {2011})}\BibitemShut {NoStop}%
\bibitem [{\citenamefont {Jolles}\ \emph {et~al.}(2017)\citenamefont {Jolles},
  \citenamefont {Boogert}, \citenamefont {Sridhar}, \citenamefont {Couzin},\
  and\ \citenamefont {Manica}}]{jolles2017consistent}%
  \BibitemOpen
  \bibfield  {author} {\bibinfo {author} {\bibfnamefont {J.~W.}\ \bibnamefont
  {Jolles}}, \bibinfo {author} {\bibfnamefont {N.~J.}\ \bibnamefont {Boogert}},
  \bibinfo {author} {\bibfnamefont {V.~H.}\ \bibnamefont {Sridhar}}, \bibinfo
  {author} {\bibfnamefont {I.~D.}\ \bibnamefont {Couzin}},\ and\ \bibinfo
  {author} {\bibfnamefont {A.}~\bibnamefont {Manica}},\ }\bibfield  {title}
  {\bibinfo {title} {Consistent individual differences drive collective
  behavior and group functioning of schooling fish},\ }\href@noop {} {\bibfield
   {journal} {\bibinfo  {journal} {Current Biology}\ }\textbf {\bibinfo
  {volume} {27}},\ \bibinfo {pages} {2862} (\bibinfo {year}
  {2017})}\BibitemShut {NoStop}%
\bibitem [{\citenamefont {Ballerini}\ \emph {et~al.}(2008)\citenamefont
  {Ballerini}, \citenamefont {Cabibbo}, \citenamefont {Candelier},
  \citenamefont {Cavagna}, \citenamefont {Cisbani}, \citenamefont {Giardina},
  \citenamefont {Lecomte}, \citenamefont {Orlandi}, \citenamefont {Parisi},
  \citenamefont {Procaccini} \emph {et~al.}}]{ballerini2008interaction}%
  \BibitemOpen
  \bibfield  {author} {\bibinfo {author} {\bibfnamefont {M.}~\bibnamefont
  {Ballerini}}, \bibinfo {author} {\bibfnamefont {N.}~\bibnamefont {Cabibbo}},
  \bibinfo {author} {\bibfnamefont {R.}~\bibnamefont {Candelier}}, \bibinfo
  {author} {\bibfnamefont {A.}~\bibnamefont {Cavagna}}, \bibinfo {author}
  {\bibfnamefont {E.}~\bibnamefont {Cisbani}}, \bibinfo {author} {\bibfnamefont
  {I.}~\bibnamefont {Giardina}}, \bibinfo {author} {\bibfnamefont
  {V.}~\bibnamefont {Lecomte}}, \bibinfo {author} {\bibfnamefont
  {A.}~\bibnamefont {Orlandi}}, \bibinfo {author} {\bibfnamefont
  {G.}~\bibnamefont {Parisi}}, \bibinfo {author} {\bibfnamefont
  {A.}~\bibnamefont {Procaccini}}, \emph {et~al.},\ }\bibfield  {title}
  {\bibinfo {title} {Interaction ruling animal collective behavior depends on
  topological rather than metric distance: Evidence from a field study},\
  }\href@noop {} {\bibfield  {journal} {\bibinfo  {journal} {Proceedings of the
  national academy of sciences}\ }\textbf {\bibinfo {volume} {105}},\ \bibinfo
  {pages} {1232} (\bibinfo {year} {2008})}\BibitemShut {NoStop}%
\bibitem [{\citenamefont {Strandburg-Peshkin}\ \emph
  {et~al.}(2013)\citenamefont {Strandburg-Peshkin}, \citenamefont {Twomey},
  \citenamefont {Bode}, \citenamefont {Kao}, \citenamefont {Katz},
  \citenamefont {Ioannou}, \citenamefont {Rosenthal}, \citenamefont {Torney},
  \citenamefont {Wu}, \citenamefont {Levin} \emph
  {et~al.}}]{strandburg2013visual}%
  \BibitemOpen
  \bibfield  {author} {\bibinfo {author} {\bibfnamefont {A.}~\bibnamefont
  {Strandburg-Peshkin}}, \bibinfo {author} {\bibfnamefont {C.~R.}\ \bibnamefont
  {Twomey}}, \bibinfo {author} {\bibfnamefont {N.~W.}\ \bibnamefont {Bode}},
  \bibinfo {author} {\bibfnamefont {A.~B.}\ \bibnamefont {Kao}}, \bibinfo
  {author} {\bibfnamefont {Y.}~\bibnamefont {Katz}}, \bibinfo {author}
  {\bibfnamefont {C.~C.}\ \bibnamefont {Ioannou}}, \bibinfo {author}
  {\bibfnamefont {S.~B.}\ \bibnamefont {Rosenthal}}, \bibinfo {author}
  {\bibfnamefont {C.~J.}\ \bibnamefont {Torney}}, \bibinfo {author}
  {\bibfnamefont {H.~S.}\ \bibnamefont {Wu}}, \bibinfo {author} {\bibfnamefont
  {S.~A.}\ \bibnamefont {Levin}}, \emph {et~al.},\ }\bibfield  {title}
  {\bibinfo {title} {Visual sensory networks and effective information transfer
  in animal groups},\ }\href@noop {} {\bibfield  {journal} {\bibinfo  {journal}
  {Current Biology}\ }\textbf {\bibinfo {volume} {23}},\ \bibinfo {pages}
  {R709} (\bibinfo {year} {2013})}\BibitemShut {NoStop}%
\bibitem [{\citenamefont {Olfati-Saber}\ and\ \citenamefont
  {Murray}(2004)}]{olfati2004consensus}%
  \BibitemOpen
  \bibfield  {author} {\bibinfo {author} {\bibfnamefont {R.}~\bibnamefont
  {Olfati-Saber}}\ and\ \bibinfo {author} {\bibfnamefont {R.~M.}\ \bibnamefont
  {Murray}},\ }\bibfield  {title} {\bibinfo {title} {Consensus problems in
  networks of agents with switching topology and time-delays},\ }\href@noop {}
  {\bibfield  {journal} {\bibinfo  {journal} {IEEE Transactions on automatic
  control}\ }\textbf {\bibinfo {volume} {49}},\ \bibinfo {pages} {1520}
  (\bibinfo {year} {2004})}\BibitemShut {NoStop}%
\bibitem [{\citenamefont {Muller}\ and\ \citenamefont
  {Stewart}(2006)}]{muller2006linear}%
  \BibitemOpen
  \bibfield  {author} {\bibinfo {author} {\bibfnamefont {K.~E.}\ \bibnamefont
  {Muller}}\ and\ \bibinfo {author} {\bibfnamefont {P.~W.}\ \bibnamefont
  {Stewart}},\ }\href@noop {} {\emph {\bibinfo {title} {Linear model theory:
  univariate, multivariate, and mixed models}}}\ (\bibinfo  {publisher} {John
  Wiley \& Sons},\ \bibinfo {year} {2006})\BibitemShut {NoStop}%
\bibitem [{\citenamefont {Kuznetsov}(1998)}]{kuznetsov1998elements}%
  \BibitemOpen
  \bibfield  {author} {\bibinfo {author} {\bibfnamefont {Y.~A.}\ \bibnamefont
  {Kuznetsov}},\ }\href@noop {} {\emph {\bibinfo {title} {Elements of applied
  bifurcation theory}}}\ (\bibinfo  {publisher} {Springer},\ \bibinfo {year}
  {1998})\BibitemShut {NoStop}%
\end{thebibliography}%

\clearpage
 \appendix
\section*{Supplementary Information}

\section{Port-Contact-Hamiltonian Formulation }\label{appendix}

Each agent evolves on a three-dimensional contact manifold $(\mathcal{M}_i,\eta_i)$ equipped with the contact one-form
\begin{equation}
\eta_i = dz_i - p_i  d\theta_i,
\end{equation}
where $\theta_i\in\mathbb{S}^1$ denotes the agent's heading, $p_i\in\mathbb{R}$ is a momentum-like activity variable, and $z_i\in\mathbb{R}$ is an internal energy- or activity-like coordinate that tracks the cumulative balance between injected and dissipated motion. The associated Reeb vector field $\mathcal{R}_i$ is uniquely defined by
\[
\iota_{\mathcal{R}_i}\eta_i = 1, 
\qquad 
\iota_{\mathcal{R}_i} d\eta_i = 0,
\]
and encodes the intrinsic dissipation direction of the contact structure.

The behavior of agent $i$ is generated by the contact Hamiltonian 
\begin{eqnarray}
K_i
&=&
\underbrace{\Big(\frac12 p_i^2-\alpha_i\cos(\theta_i-\phi_i)\Big)}_{\text{local conservative storage}}\\
&+& 
\underbrace{\Big(-\kappa\sum_j a_{ij}\cos(\theta_j-\theta_i)\Big)}_{\text{network interaction}}
\\ &+&  
\underbrace{\gamma_i z_i}_{\text{contact/dissipative part}}
\label{eq:contact-Hamiltonian-SI}
\end{eqnarray}
where $\alpha_i\ge 0$ encodes the strength of directional preference toward a target direction $\phi_i$, $\kappa>0$ is the alignment gain, $A=[a_{ij}]$ is the weighted adjacency matrix, and $\gamma_i>0$ is a dissipation coefficient.
The last term makes dissipation intrinsic to the geometry, rather than externally imposed.
The contact Hamiltonian vector field $X_{K_i}$ is defined by
\begin{equation}
\iota_{X_{K_i}} d\eta_i = dK_i - (\mathcal{R}_i K_i) \eta_i,
\qquad 
\iota_{X_{K_i}}\eta_i = -K_i.
\end{equation}
Writing $X_{K_i}=a_i\partial_{\theta_i}+b_i\partial_{p_i}+c_i\partial_{z_i}$ yields the coordinate equations
\begin{align}
\dot{\theta}_i &= \partial_{p_i} K_i, \\
\dot{p}_i &= -\partial_{\theta_i} K_i - p_i\partial_{z_i}K_i,\\
\dot{z}_i &= p_i\partial_{p_i}K_i - K_i.
\end{align}
Substituting~\eqref{eq:contact-Hamiltonian-SI} gives the agent-level dynamics governed by the equations of motion (EoM)
\begin{align}
\dot{\theta}_i &= p_i, \label{eq:model-theta}\\
\dot p_i &= -\alpha_i\sin(\theta_i - \phi_i)
           +\kappa \sum_{j=1}^{N} a_{ij}\sin(\theta_j - \theta_i)
           -\gamma_i p_i, \label{eq:model-p}\\
\dot z_i &= \tfrac12 p_i^2 
           + \alpha_i\cos(\theta_i-\phi_i)
           + \kappa \sum_{j=1}^{N} a_{ij}\cos(\theta_j - \theta_i)
           - \gamma_i z_i. \label{eq:model-z}
\end{align}

The variable $z_i$ is not introduced to create an additional observable decision coordinate. Rather, it provides the minimal contact-geometric state extension through which dissipation is represented intrinsically. In this sense, $z_i$ acts as a bookkeeping variable that records the cumulative balance between activity injected by leader forcing and alignment, and activity removed by local damping. This is what allows the dissipative mechanism to be encoded within the geometry itself and yields a transparent network-level energy-dissipation identity. Importantly, $z_i$ does not feed back into the heading dynamics; it serves to separate conservative redistribution of activity from irreversible loss. Without $z_i$, the same reduced dynamics for $(\theta_i,p_i)$ can still be written, but the dissipative mechanism is no longer internal to the geometry, and the associated energetic interpretation becomes far less transparent. In particular, the equation for $\dot{z}_i$ shows how alignment and leader forcing inject activity into the internal coordinate, while the term $-\gamma_i z_i$ causes this activity to dissipate unless continuously sustained. The damping term $-\gamma_i p_i$ acts on the rate of reorientation rather than through a preferred target direction. In this sense, damping modifies persistence without injecting additional directional preference.

Notably, for small heading differences, the alignment term in~\eqref{eq:model-p} linearizes to a Laplacian smoothing term
\[
\sum_{j=1}^{N} a_{ij}\sin(\theta_j-\theta_i) \approx \sum_{j=1}^{N} a_{ij}(\theta_j-\theta_i)
= -(L\boldsymbol{\theta})_i,
\]
so \(L\) governs the relaxation of perturbations away from consensus.
In the second-order dynamics, these perturbations evolve as damped network modes, making the Fiedler direction the natural carrier of the slow instability that organizes the compromise-to-decision transition. 
In structured interaction graphs, the extent to which this mode can coordinate the full group may additionally depend on sparse bridges or cut-sets, which can restrict the transmission of alignment between modules.

\section{Global energy-dissipation}

To expose the network interconnection structure, define the per-agent stored energy $H_i$
\[
H_i(\theta_i,p_i)
=
\tfrac12 p_i^2
- \alpha_i \cos(\theta_i-\phi_i),
\]
which is related to the global conservative storage via
\[H = \sum_{i=1}^N H_i -\frac{\kappa}{2}\sum_{i,j} a_{ij}\cos(\theta_j-\theta_i).
\]
Evaluating the time derivative of $H_i$ along trajectories of the system gives
\begin{align*}
\dot{H}_i
&=
p_i\dot{p}_i + \alpha_i\sin(\theta_i-\phi_i)\dot{\theta}_i
\\
&=
\kappa p_i\sum_{j=1}^N a_{ij}\sin(\theta_j-\theta_i)
-\gamma_i p_i^2.
\end{align*}

Introducing the port variables 
\[
u_i := \kappa \sum_{j=1}^N a_{ij}\sin(\theta_j-\theta_i),
\qquad 
y_i := p_i,
\]
we obtain the nodal passivity identity
\begin{equation}
\dot{H}_i = u_i y_i - \gamma_i y_i^2.
\label{eq:nodal-passivity}
\end{equation}
Thus, each agent is passive from input $u_i$ (social alignment forces) to output $y_i$ (rate of heading change), with internal dissipation governed by $\gamma_i$.

For undirected networks, the alignment interconnection is conservative. Accordingly, once the pairwise interaction energy is included in the global conservative storage \(H\), the interconnection contribution cancels in the total energy. 
The corresponding global energy-dissipation identity (proved below in Theorem~A) for the conservative storage \(H\) is

\begin{equation}
\label{dotH}
\dot H
=
-\sum_{i=1}^N \gamma_i p_i^2
\le 0,
\end{equation}

Notably, alignment interactions redistribute energy across agents without generating it, while dissipation - particularly from uninformed individuals with $\alpha_i\simeq 0$ and $\gamma_i>0$ - removes energy irreversibly.
This structural separation between conservative redistribution and irreversible loss is precisely what underlies the energy-dissipation balance of the network.

This structural property underlies the stability results in the main text; dissipation constrains the growth of unstable collective modes without introducing directional bias. In structured networks, whether this stabilization is sufficient at the group level may also depend on how effectively alignment can pass through sparse bridges or cut-sets connecting different modules. In particular, uninformed individuals, characterized by $\alpha_i \simeq 0$ and $\gamma_i > 0$, do not contribute preferred target directions, but they do contribute dissipation that acts through the same collective dynamics and therefore enhances stabilization of compromise.

The proof below shows explicitly that the global identity~\eqref{dotH} follows from the symmetry of the interaction term together with the fact that the contact contribution enters only through local dissipation.

\subsection*{Lemma A} ~Let $S=\frac{\kappa}{2}\sum_{i,j} a_{ij}(p_i-p_j)\sin(\theta_j-\theta_i)$, if  $a_{ij}=a_{ji}$ then
$
S=\kappa\sum_i p_i\sum_j a_{ij}\sin(\theta_j-\theta_i)$.

\vspace{2mm}
\noindent\smallbreak \textit{Proof.} Expand $S$ to obtain
\[
S
=
\frac{\kappa}{2}\sum_{i,j} a_{ij}p_i\sin(\theta_j-\theta_i)
-
\frac{\kappa}{2}\sum_{i,j} a_{ij}p_j\sin(\theta_j-\theta_i).
\]
In the second term, relabel \(i\leftrightarrow j\), then using that \(a_{ji}=a_{ij}\) and
\(\sin(\theta_i-\theta_j)=-\sin(\theta_j-\theta_i)\), gives the result.

\subsection*{Theorem A} 
 For the contact Hamiltonian
\beq
K_i
&=
\left(\frac12 p_i^2-\alpha_i\cos(\theta_i-\phi_i)\right)\\
 &+ 
\left(-\kappa\sum_j a_{ij}\cos(\theta_j-\theta_i)\right)
 + 
\gamma_i z_i.
\label{eq:Ki-split}
\eeq
where the network interaction is symmetric \(a_{ij}=a_{ji}\), the global conservative storage function $H$ satisfies
\[
\dot H
=
-\sum_{i=1}^N \gamma_i p_i^2\le 0.
\]

\vspace{3mm}

\noindent\textit{Proof.}
For a given contact Hamiltonian \(K_i\)  the function \(H(\boldsymbol\theta,\boldsymbol p)\) is constructed from \(K_i\) by:
\begin{enumerate}
\item Removing the contact term \(\gamma_i z_i\), since it is not part of the conservative
\((\theta,p)\)-energy;
\item Summing the remaining \((\theta_i,p_i)\)-dependent contributions over all nodes;
\item A factor \(1/2\) is included in the interaction term to avoid double-counting edges.
\end{enumerate}
Under this prescription, it follows that
\[
H(\boldsymbol\theta,\boldsymbol p)
=
\sum_{i=1}^N\left(\frac12 p_i^2-\alpha_i\cos(\theta_i-\phi_i)\right)
-\frac{\kappa}{2}\sum_{i,j} a_{ij}\cos(\theta_j-\theta_i).
\]
We then differentiate \(H\) which yields the intermediate result
\begin{align*}
\dot H
&=
\sum_{i=1}^N\left[
p_i\dot p_i+\alpha_i\sin(\theta_i-\phi_i)\dot\theta_i
\right]
-\frac{\kappa}{2}\frac{d}{dt}\sum_{i,j} a_{ij}\cos(\theta_j-\theta_i).
\end{align*}
For the interaction term we have that
\begin{equation}
\frac{d}{dt}\cos(\theta_j-\theta_i)
=
-\sin(\theta_j-\theta_i)(\dot\theta_j-\dot\theta_i)
=
(p_i-p_j)\sin(\theta_j-\theta_i),\label{2}
\end{equation}
Moreover, from the contact Hamiltonian the equations of motion are
\[
\dot\theta_i=p_i,
\qquad
\dot p_i=
-\alpha_i\sin(\theta_i-\phi_i)
+\kappa\sum_{j=1}^N a_{ij}\sin(\theta_j-\theta_i)
-\gamma_i p_i.
\]
Substituting \(\dot\theta_i=p_i\) and \(\dot p_i\)  into $\dot H$ and using eq.~(\ref{2}) one obtains
\begin{eqnarray}
\dot H
&=&
-\sum_i p_i\alpha_i\sin(\theta_i-\phi_i)
+\kappa\sum_i p_i\sum_j a_{ij}\sin(\theta_j-\theta_i)\nonumber\\
&-&\sum_i \gamma_i p_i^2
\qquad
+\sum_i \alpha_i\sin(\theta_i-\phi_i)p_i\nonumber\\
&-&\frac{\kappa}{2}\sum_{i,j} a_{ij}(p_i-p_j)\sin(\theta_j-\theta_i).\nonumber
\end{eqnarray}
By definition,
$S=\frac{\kappa}{2}\sum_{i,j} a_{ij}(p_i-p_j)\sin(\theta_j-\theta_i)$, and by lemma A,
$
S=\kappa\sum_i p_i\sum_j a_{ij}\sin(\theta_j-\theta_i).$
Thus the interaction terms cancel, as do the \(\alpha_i\)-terms, leaving
\[
\dot H
=
-\sum_{i=1}^N \gamma_i p_i^2\le 0
\]
which is negative semidefinite since $\gamma_i>0$ and vanishes only when $p_i=0$ for all $i$.
This proves the exact identity stated in~\eqref{dotH}, and in particular shows that all irreversible loss in the network arises solely from the local damping terms.

\section{Interpretation of model components and parameters}

The model presented in the main text provides a geometric Hamiltonian-inspired description of collective decision dynamics relevant to biological and engineering systems. In this section we provide a physical interpretation of each of the relevant quantities intrinsic to the model.
The value of this interpretation is not to replace connectivity-based accounts of collective decision-making~\citep{couzin2005effective,leonard2012decision,olfati2007consensus}, but to clarify which additional aspects become analyzable once response persistence and local dissipation are included explicitly. In particular, the model links parameters not only to collective outcomes, but also to transient response, modal stability, and experimentally or algorithmically tunable quantities.

The heading \(\theta_i\in\mathbb{S}^1\) is the observable decision variable, representing the expressed direction of motion. The variable \(p_i\in\mathbb{R}\) captures persistence of reorientation. It provides a coarse-grained description of how rapidly an individual turns in response to social and directional cues. This is consistent with the fact that biological agents exhibit smooth reorientation over finite time, rather than instantaneous heading jumps, due to biomechanical and behavioral constraints~\citep{Cavagna2014bird,attanasi2014collective}. 
The bookkeeping variable $z_i \in \mathbb{R}$ should be interpreted as an internal contact-geometric coordinate rather than as a directly measurable biological state. Its role is to encode, within the geometric formulation, the cumulative balance between activity injected by alignment and leader forcing and activity removed by local damping. In this sense, $z_i$ provides the minimal state extension needed to represent dissipation intrinsically and to recover a closed network-level energy-dissipation balance.

The coupling strength \(\kappa\) scales social responsiveness. In animal groups, it may summarize effective sensitivity to neighbors, including perception, attention, and behavioral gain~\citep{herbert2011inferring,katz2011inferring}; in robotic swarms, it plays the role of an alignment gain relating observed neighbor headings to turning commands~\citep{olfati2007consensus}. The damping coefficient \(\gamma_i\) sets how quickly reorientation activity decays. Biologically, it can be interpreted as loss of persistence or resistance to sustained turning under weak or conflicting cues~\citep{Cavagna2014bird,attanasi2014collective}, in engineered systems, it corresponds to local damping or control dissipation~\citep{olfati2007consensus}. Heterogeneity in \(\gamma_i\) provides a simple way to encode differences in manoeuvrability or behavioral type across individuals, which are known to affect collective outcomes~\citep{jolles2017consistent}.
Because the damping acts on the rate of reorientation rather than through an additional preferred target direction, it modifies persistence without injecting further directional bias into the decision dynamics.

The parameter \(\alpha_i\ge 0\) quantifies informedness by setting the strength of directional preference toward a target \(\phi_i\). Thus, informed leaders have \(\alpha_i>0\), whereas uninformed agents correspond to \(\alpha_i\approx 0\). These uninformed agents contribute no directional bias but remain dynamically relevant through alignment and damping~\citep{couzin2005effective,leonard2012decision}. Finally, the weighted adjacency matrix \(A=(a_{ij})\) encodes who interacts with whom. In animal groups, effective interactions are often better described as topological or weighted-neighbor interactions than as purely metric ones~\citep{ballerini2008interaction,katz2011inferring,herbert2011inferring}, while sensory constraints such as anisotropy or occlusion can induce time-varying interaction structure~\citep{strandburg2013visual}. At a coarse-grained level, the Laplacian eigenvalue \(\lambda_2\) summarizes the strength of the weakest non-consensus coordination mode, making it a standard indicator of network-level coordination performance~\citep{olfati2004consensus,olfati2007consensus}. 
In modular or weakly connected graphs, however, coordination may also be limited by sparse bridges or cut-sets, so that the ability of this slow mode to synchronize the full group depends not only on its eigenvalue but also on how alignment is transmitted across topological bottlenecks.

\section{Main Theorems}

In what follows we provide the proofs of the main theorems stated in the main text. We begin with the low-conflict stable-compromise result, which combines existence and local exponential stability in a single statement.

\subsection*{Theorem 1 (Stable compromise)}
\emph{Assume the interaction graph is undirected and connected, and that at least one agent is informed, i.e.\ $\alpha_i>0$ for some $i$. If the conflict $\Delta$ is sufficiently small, then the system admits a locally unique compromise equilibrium $(\tilde{\boldsymbol\theta}^\ast(\Delta),\boldsymbol p^\ast)$ near $(0,0)$, with $\boldsymbol p^\ast=0$. Further, this equilibrium is locally exponentially stable.}

\medskip

\noindent\textit{Proof.}
We divide the proof into two steps: existence of the compromise equilibrium, and its local exponential stability.

\medskip

\noindent\textbf{Step 1: Existence of the compromise equilibrium.}
We work in the co-moving frame $\tilde\theta_i=\theta_i-\bar\phi,$ where \(\bar\phi\) is the average preferred direction of the informed leaders, and write
\[
\tilde\phi_i(\Delta):=\phi_i-\bar\phi.
\]
In this frame the equations of motion are
\[
\dot{\tilde\theta}_i = p_i,
\,
\dot p_i=
-\alpha_i\sin(\tilde\theta_i-\tilde\phi_i(\Delta))
+\kappa\sum_{j=1}^N a_{ij}\sin(\tilde\theta_j-\tilde\theta_i)
-\gamma_i p_i.
\]

An equilibrium is a time-independent solution, so it must satisfy
$\dot{\tilde\theta}_i=0$, and
$\dot p_i=0$.
Since \(\dot{\tilde\theta}_i=p_i\), the first condition immediately implies for all \(i\)
\[
p_i^\ast=0
\]
Substituting \(p_i^\ast=0\) into the second equation and imposing \(\dot p_i=0\) gives the equilibrium condition
\[
0=
-\alpha_i\sin(\tilde\theta_i^\ast-\tilde\phi_i(\Delta))
+\kappa\sum_{j=1}^N a_{ij}\sin(\tilde\theta_j^\ast-\tilde\theta_i^\ast).
\]
It is convenient to define
\begin{equation}
\label{Fi}
F_i(\tilde{\boldsymbol\theta},\Delta)
:=
-\alpha_i\sin(\tilde\theta_i-\tilde\phi_i(\Delta))
+\kappa\sum_{j=1}^N a_{ij}\sin(\tilde\theta_j-\tilde\theta_i),
\end{equation}
so that equilibria are precisely the solutions of $F(\tilde{\boldsymbol\theta},\Delta)=0.$

At \(\Delta=0\), all preferred directions coincide in the co-moving frame, so \(\tilde\phi_i(0)=0\) for all informed agents. Hence
$\tilde{\boldsymbol\theta}=0$
is an equilibrium configuration.
The implicit function theorem states that if \(F:\mathbb R^N\times\mathbb R\to\mathbb R^N\) is smooth, \(F(0,0)=0\), and the Jacobian \(D_{\tilde{\boldsymbol\theta}}F(0,0)\) is invertible, then there exists a neighborhood of \(\Delta=0\) and a unique smooth function \(\tilde{\boldsymbol\theta}^\ast(\Delta)\) near \(0\) such that
\begin{equation}
\label{Fi0}
F(\tilde{\boldsymbol\theta}^\ast(\Delta),\Delta)=0
\end{equation}
for all sufficiently small \(\Delta\).

Thus to determine whether this equilibrium persists for small \(\Delta\), it remains to compute the Jacobian of \(F\) with respect to \(\tilde{\boldsymbol\theta}\) at \((\tilde{\boldsymbol\theta},\Delta)=(0,0)\); that is
\[
J_0 := \left. \frac{\partial F_i}{\partial \tilde\theta_k} \right|_{(0,0)}=-\bigl(\mathrm{diag}(\boldsymbol\alpha)+\kappa L\bigr),
\]
where the latter expression follows from direct calculation and involves the graph Laplacian \(L\).

We now show that \(J_0\) is nonsingular on the full space \(\mathbb R^N\). For any nonzero vector \(x\in\mathbb R^N\),
\[
\langle x,-J_0x\rangle
=
x^\top\bigl(\mathrm{diag}(\boldsymbol\alpha)+\kappa L\bigr)x
=
\sum_{i=1}^N \alpha_i x_i^2+\kappa x^\top Lx.
\]
Because the graph is undirected and connected, the Laplacian \(L\) is symmetric and positive semidefinite, so
$x^\top Lx\ge 0,$ with equality if and only if \(x=c\mathbf{1}\) for some constant \(c\). If \(x\) is not proportional to \(\mathbf 1\), then \(x^\top Lx>0\), and therefore
\[
\langle x,-J_0x\rangle>0.
\]
If instead \(x=c\mathbf 1\) with \(c\neq 0\), then
\[
\langle x,-J_0x\rangle=\sum_{i=1}^N \alpha_i x_i^2
=
c^2\sum_{i=1}^N \alpha_i
>0,
\]
since at least one \(\alpha_i>0\). Hence in all cases $\langle x,-J_0x\rangle>0$ for $x\neq 0$. Thus \(-J_0\) is positive definite, and \(J_0\) is invertible.

Having established the existence of the compromise equilibrium, we now evaluate its linear stability.

\medskip

\noindent\textbf{Step 2: Local exponential stability.}
Consider a general perturbation around the equilibrium point \((\tilde{\boldsymbol\theta}^\ast(\Delta),0)\) given by
\[
\tilde{\boldsymbol\theta}
=
\tilde{\boldsymbol\theta}^\ast(\Delta)+\boldsymbol\eta,
\qquad
\boldsymbol p=\boldsymbol\nu.
\]
Since
$\dot{\tilde\theta}_i=p_i$,
it follows that
\[
\dot{\boldsymbol\eta}=\boldsymbol\nu.
\]
Moreover, the momentum equation of motion can be expressed as
\begin{equation}\label{p1}
\dot{\boldsymbol p}=F(\tilde{\boldsymbol\theta};\Delta)-\Gamma \boldsymbol p,
\end{equation}
where $\Gamma:=\mathrm{diag}(\gamma_i)\succ 0$ and $F$ is defined above.

Since $\tilde{\boldsymbol\theta}^\ast(\Delta)$ is an equilibrium,
$F(\tilde{\boldsymbol\theta}^\ast(\Delta),\Delta)=0$ and a first-order Taylor expansion of $F$ around
$\tilde{\boldsymbol\theta}^\ast(\Delta)$ gives
\[
F(\tilde{\boldsymbol\theta}^\ast(\Delta)+\boldsymbol\eta,\Delta)
=
F(\tilde{\boldsymbol\theta}^\ast(\Delta),\Delta)
+
J(\Delta)\boldsymbol\eta
+
\mathcal O(\|\boldsymbol\eta\|^2),
\]
in terms of the Jacobian of $F$ evaluated at the equilibrium
\[
J(\Delta):=
D_{\tilde{\boldsymbol\theta}}F(\tilde{\boldsymbol\theta}^\ast(\Delta),\Delta).
\]

Using $F(\tilde{\boldsymbol\theta}^\ast(\Delta),\Delta)=0$, at leading order
\[
F(\tilde{\boldsymbol\theta}^\ast(\Delta)+\boldsymbol\eta,\Delta)
=
J(\Delta)\boldsymbol\eta.
\]
Substituting into eq.~(\ref{p1}) gives
\[
\dot{\boldsymbol\nu}=J(\Delta)\boldsymbol\eta-\Gamma\boldsymbol\nu.
\]
Hence the linearized system is
\[
\dot{\boldsymbol\eta}=\boldsymbol\nu,
\qquad
\dot{\boldsymbol\nu}=J(\Delta)\boldsymbol\eta-\Gamma\boldsymbol\nu.
\]
Eliminating \(\boldsymbol\nu\), this is equivalent to the second-order system
\begin{equation}
\ddot{\boldsymbol\eta}+\Gamma\dot{\boldsymbol\eta}+K\boldsymbol\eta=0,
\label{lin}
\end{equation}
with \(K:=-J(\Delta)\).
Recall from Step 1 that \(K:=-J_0\succ 0\). By continuity of \(J(\Delta)\) along the equilibrium branch, since
\[
J_0=-(\mathrm{diag}(\boldsymbol\alpha)+\kappa L),
\]
the matrix \(J(\Delta)\) remains negative definite for sufficiently small \(\Delta\).

Thus the linearized system is a damped second-order system with a positive-definite stiffness matrix \(-J(\Delta)\) and a positive-definite damping matrix \(\Gamma\).
Equivalently, the linearization matrix
\[
A(\Delta):=
\begin{pmatrix}
0&I\\
J(\Delta)&-\Gamma
\end{pmatrix}
\]
has no spectrum on the imaginary axis when \(J(\Delta)\prec 0\) and \(\Gamma\succ 0\).
Standard results for damped second-order systems with positive-definite stiffness and damping matrices imply that all solutions decay exponentially to \(0\) \ \cite{muller2006linear}. Hence the equilibrium is locally exponentially stable.
\hfill\(\square\)

{\bf Remark 1.}
As \(\Delta\) increases, there will typically exist a finite critical value \(\Delta=\Delta_c\) such that for $\Delta>\Delta_c$ the Jacobian \(J(\Delta)\) ceases to be negative definite. This marks the onset of linear instability of the compromise equilibrium.

{\bf Remark 2.}
The small-conflict condition should be understood relative to the effective restoring scale of the networked system. In particular, larger values of \(\kappa\lambda_2\) increase the stiffness of the weakest non-consensus mode and therefore enlarge the range of \(\Delta\) over which \(J(\Delta)\) remains negative definite. Indeed, since
\[
x^\top\bigl(\mathrm{diag}(\boldsymbol\alpha)+\kappa L\bigr)x
=
\sum_{i=1}^N \alpha_i x_i^2+\kappa x^\top Lx,
\]
the contribution of the Laplacian is controlled from below on the non-consensus subspace by \(\kappa\lambda_2\).
Thus, stronger coupling or better-connected interaction graphs make the compromise equilibrium more robust to increasing $\Delta$.

\medskip 

In the modular regime considered here, increasing conflict loads the sparse inter-module cut-set more heavily while intra-module deformations remain small. Theorem~2 makes this precise through a rigid-module reduction in which the leading instability is an inter-module contrast mode, and the compromise branch is lost when the reduced two-module equilibrium system reaches a nondegenerate fold.

\medskip


\subsection*{Theorem 2. (Bottleneck-mediated fold of compromise)}
\emph{
Consider a modular network consisting of two connected modules \(M_A\) and \(M_B\), coupled by a sparse cut-set \(C\) of cross-module pairs \((i,j)\) with \(i\in M_A\), \(j\in M_B\), each edge counted once, with leaders in \(M_s\) preferring target \(\phi_s\), \(s\in\{A,B\}\), and conflict angle \(\Delta=|\phi_A-\phi_B|\).
Let
$
\alpha^s := \sum_{i\in M_s}\alpha_i$, and
\[
K_C := \kappa\sum_{(i,j)\in C} a_{ij},
\qquad
\eta := \frac{\max(\alpha^A,\alpha^B,K_C)}{\kappa\min_s \lambda_2(L_s)}.
\]
For \(\eta\ll1\) each module remains internally rigid to leading order. Suppose that as \(\Delta\) varies in \((0,\pi)\), the corresponding rigid-module reduced system admits a compromise branch that undergoes a nondegenerate saddle-node bifurcation at some
\(
\Delta_{\mathrm{fold}}^{(0)}\in(0,\pi),
\)
 Further, assume that the reduced compromise branch stays in the regime where each module is on the restoring side of its leader forcing.
Then, for \(\eta\) sufficiently small,
the compromise branch persists for \(\Delta\in[0,\Delta_{\mathrm{fold}})\) and terminates at \(\Delta=\Delta_{\mathrm{fold}}\in(0,\pi)\) through a saddle-node bifurcation. Moreover, the associated critical mode is an inter-module contrast mode, concentrated on the bottleneck transmission direction, and has the form
\[
\varphi = c_A \mathbf{1}_A + c_B \mathbf{1}_B + O(\eta),
\qquad c_A c_B<0.
\]
}

\medskip

\noindent\textit{Proof.}
The proof has four steps. We first derive the rigid-module reduced equilibrium equations, then show that this reduced system develops a fold in the relevant parameter regime, next lift that fold back to the full network for \(\eta\) sufficiently small, and finally identify the corresponding critical mode in the full Jacobian.

\medskip
\noindent\textbf{Step 1: Rigid-module reduction.}
For \(i\in M_s\), write
\[
\tilde\theta_i=\Theta_s+\chi_i,
\qquad
\sum_{i\in M_s}\chi_i=0,
\qquad s\in\{A,B\},
\]
where
\[
\Theta_s:=\frac{1}{|M_s|}\sum_{i\in M_s}\tilde\theta_i
\]
is the mean phase of module \(M_s\), and \(\chi_i\) measures the intra-module fluctuation about that mean. Thus the zero-mean subspace of module \(M_s\) is
\[
\chi_s:=(\chi_i)_{i\in M_s}\in \mathcal X_s,
\qquad
\mathcal X_s:=\Bigl\{x\in\mathbb R^{|M_s|}:\sum_{i\in M_s}x_i=0\Bigr\}.
\]

By assumption
\(\eta\ll 1,\)
so the intra-module restoring scale \(\kappa\lambda_2(L_s)\) dominates both the leader forcing scale and the inter-module bridge scale.
We now project the equilibrium equations
\(
F_i(\tilde{\boldsymbol\theta},\Delta)=0,
\)
with \(F_i\) defined in eq.~\eqref{Fi}, onto the zero-mean subspace \(\mathcal X_s\). 
Writing \(P_s\) for the orthogonal projection onto \(\mathcal X_s\), and separating intra-module from inter-module couplings, the equilibrium equations on module \(M_s\) may be written in vector form as
\begin{widetext}
\begin{equation}
0=
P_s \left[
\left(
-\alpha_i\sin(\Theta_s+\chi_i-\tilde\phi_i)
+\kappa\sum_{j\in M_s}a_{ij}\sin(\chi_j-\chi_i)
+\kappa\sum_{j\notin M_s}a_{ij}\sin(\tilde\theta_j-\tilde\theta_i)
\right)_{i\in M_s}
\right].
\end{equation}
\end{widetext}
Since \(\chi_j-\chi_i\) is small in the rigid-module regime, the intra-module sine term linearizes as
\[
\sin(\chi_j-\chi_i)=(\chi_j-\chi_i)+O(|\chi_j-\chi_i|^3),
\]
and therefore
\[
\kappa\sum_{j\in M_s}a_{ij}\sin(\chi_j-\chi_i)
=
-(\kappa L_s\chi_s)_i+\mathcal N_{s,i}(\chi_s),
\]
where \(\mathcal N_s(\chi_s)=O(\|\chi_s\|^2)\) for \(\chi_s\) sufficiently small. Hence the projected fluctuation equation may be written as
\[
\kappa L_s\chi_s
=
\mathcal R_s(\Theta_A,\Theta_B,\chi,\Delta),
\]
where \(\mathcal R_s\) collects the projected leader forcing, the inter-module coupling terms, and the higher-order remainder \(\mathcal N_s\).

The key point is that \(\mathcal R_s\) is forced only by the leader strengths and the cut-set coupling. More precisely, on a fixed neighborhood of the compromise branch,
\[
\|\mathcal R_s\|
\le
C\Bigl(\max(\alpha^A,\alpha^B,K_C)+\|\chi_s\|^2\Bigr),
\]
for some constant \(C\) independent of \(\eta\). Since \(L_s\) is the Laplacian of a connected module, its kernel is spanned by the constant vector \(\mathbf 1_s\), and its restriction to \(\mathcal X_s=\mathbf 1_s^\perp\) is invertible. Moreover, because \(L_s\) is symmetric and positive definite on \(\mathcal X_s\),
\[
\bigl\|(L_s|_{\mathcal X_s})^{-1}\bigr\|
\le
\lambda_2(L_s)^{-1}.
\]
Applying \((L_s|_{\mathcal X_s})^{-1}\) to the fluctuation equation therefore gives
\[
\|\chi_s\|
\le
\frac{C}{\kappa\lambda_2(L_s)}
\Bigl(\max(\alpha^A,\alpha^B,K_C)+\|\chi_s\|^2\Bigr).
\]
Rearranging gives
\[
\|\chi_s\|
\left(
1-\frac{C\|\chi_s\|}{\kappa\lambda_2(L_s)}
\right)
\le
\frac{C \max(\alpha^A,\alpha^B,K_C)}{\kappa\lambda_2(L_s)}.
\]
At \(\eta=0\), the right-hand side vanishes and the fluctuation equation has the unique solution \(\chi_s=0\). Hence, by continuity of the local solution branch provided by the Implicit Function Theorem, \(\chi_s\) remains in a sufficiently small neighborhood of \(0\) for all sufficiently small \(\eta\). In particular, for \(\eta\) small enough one has
\[
\frac{C\|\chi_s\|}{\kappa\lambda_2(L_s)}\le \frac12,
\]
so the factor in parentheses is bounded below by \(1/2\). It follows that
\[
\|\chi_s\|
\le
\frac{2C \max(\alpha^A,\alpha^B,K_C)}{\kappa\lambda_2(L_s)}
=
O(\eta).
\]
Since all norms are equivalent in finite dimension, this implies the componentwise bound
\(\max_{i\in M_s}|\chi_i|=O(\eta),\)
and hence
\[
\max_{i,j\in M_s}|\tilde\theta_i-\tilde\theta_j|
=
O(\eta).
\]
Thus each module is internally rigid to leading order.

We now derive the reduced equations for the module means. Summing
\[
F_i(\tilde{\boldsymbol\theta},\Delta)=0
\]
over \(i\in M_A\) and using symmetry \(a_{ij}=a_{ji}\), the intra-module alignment terms cancel pairwise:
\[
\sum_{i\in M_A}\sum_{j\in M_A}a_{ij}\sin(\tilde\theta_j-\tilde\theta_i)=0.
\]
Thus only leader forcing and cut-set coupling remain. Using
\[
\tilde\theta_i=\Theta_A+O(\eta)\quad(i\in M_A),
\qquad
\tilde\theta_j=\Theta_B+O(\eta)\quad(j\in M_B),
\]
together with the assumption that all leaders in \(M_A\) share the same target \(\phi_A\), we obtain
\[
\sum_{i\in M_A}\alpha_i\sin(\tilde\phi_A-\tilde\theta_i)
=
\alpha^A\sin(\tilde\phi_A-\Theta_A)+O(\eta),
\]
and
\[
\kappa\sum_{i\in M_A}\sum_{j\in M_B}a_{ij}\sin(\tilde\theta_j-\tilde\theta_i)
=
K_C\sin(\Theta_B-\Theta_A)+O(\eta).
\]
Hence the module-sum equation reduces to
\[
\alpha^A\sin(\tilde\phi_A-\Theta_A)+K_C\sin(\Theta_B-\Theta_A)=O(\eta).
\]
The same argument on \(M_B\) gives
\[
\alpha^B\sin(\tilde\phi_B-\Theta_B)+K_C\sin(\Theta_A-\Theta_B)=O(\eta).
\]
Therefore the leading-order equilibrium equations for the rigid-module reduced system are (up to  \(O(\eta)\))
\begin{equation}
\label{eq:reducedR}
\begin{aligned}
\alpha^A\sin(\tilde\phi_A-\Theta_A)+K_C\sin(\Theta_B-\Theta_A) &= 0,\\
\alpha^B\sin(\tilde\phi_B-\Theta_B)+K_C\sin(\Theta_A-\Theta_B) &= 0.
\end{aligned}
\end{equation}

\medskip
\noindent\textbf{Step 2: Fold of the reduced system.}
Define the bridge phase difference
\[
\delta:=\Theta_B-\Theta_A.
\]
At \(\Delta=0\), the reduced system~\eqref{eq:reducedR} has the symmetric compromise equilibrium
\(\Theta_A=\Theta_B=0,\)
and its Jacobian at that point is nonsingular. Hence, by the Implicit Function Theorem, for sufficiently small \(\Delta\) there exists a smooth local compromise branch of the reduced system near \(\Theta_A=\Theta_B=0\).

By assumption, as \(\Delta\) varies in \((0,\pi)\), the reduced compromise branch of~\eqref{eq:reducedR} undergoes a nondegenerate saddle-node bifurcation at some
\[
\Delta_{\mathrm{fold}}^{(0)}\in(0,\pi),
\]
which we take to be the smallest such value.
Equivalently, if
\(
G(\Theta_A,\Theta_B,\Delta)
\)
denotes the vector field of the reduced system~\eqref{eq:reducedR}, then at the fold point
\[
G(\Theta_A^c,\Theta_B^c,\Delta_{\mathrm{fold}}^{(0)})=0,
\]
and the Jacobian
\[
J_{\mathrm{red}}^c
:=
D_{(\Theta_A,\Theta_B)}G(\Theta_A^c,\Theta_B^c,\Delta_{\mathrm{fold}}^{(0)})
\]
has a simple zero eigenvalue.
Let \(v\) and \(w\) denote corresponding right and left null vectors, normalized by \(w^\top v=1\). By the standard Lyapunov-Schmidt characterization of a saddle-node bifurcation \cite{kuznetsov1998elements}, the reduced scalar equation has the local form
\[
g(\xi,\mu)=a \mu+b \xi^2+\cdots,
\qquad \mu:=\Delta-\Delta_{\mathrm{fold}}^{(0)},
\]
with \cite{kuznetsov1998elements}
\begin{equation}
\begin{aligned}
a &=
w^\top \partial_\Delta G(\Theta_A^c,\Theta_B^c,\Delta_{\mathrm{fold}}^{(0)}),
\\
b &=
\frac12 w^\top D^2_{(\Theta_A,\Theta_B)}G(\Theta_A^c,\Theta_B^c,\Delta_{\mathrm{fold}}^{(0)})(v,v).
\end{aligned}
\end{equation}
The fold is nondegenerate precisely when
\[
a\neq 0
\qquad\text{and}\qquad
b\neq 0,
\]
corresponding respectively to transversality and nonvanishing curvature.

\medskip
\noindent\textbf{Step 3: Persistence of the fold in the full system.}
The full equilibrium equations, written in the decomposition
\[
\tilde\theta_i=\Theta_s+\chi_i,
\]
take the form
\[
\mathcal F(\Theta_A,\Theta_B,\chi,\Delta;\eta)=0.
\]
We split these equations into their zero-mean fluctuation part and their module-mean part. For fixed \((\Theta_A,\Theta_B,\Delta)\) near the reduced fold point, the fluctuation equations act on the zero-mean subspaces of the two modules. Their linearization with respect to \(\chi\) is given at leading order by the block operator \(\kappa \mathrm{diag}(L_A,L_B)\), which is invertible on the zero-mean subspaces because each \(L_s\) has spectral gap \(\lambda_2(L_s)>0\). The remaining terms are \(O(\max(\alpha^A,\alpha^B,K_C))\), hence are perturbative when \(\eta\ll1\).

Therefore, by the Implicit Function Theorem, for \(\eta\) sufficiently small there exists a unique smooth map
\[
\chi = X(\Theta_A,\Theta_B,\Delta;\eta),
\qquad
X=O(\eta),
\]
solving the fluctuation equations locally. Substituting this relation into the module-mean equations yields an effective reduced equilibrium system
\[
G_\eta(\Theta_A,\Theta_B,\Delta)=0.
\]
Because \(X=O(\eta)\) and the full equilibrium equations are smooth, the effective reduced vector field satisfies
\[
G_\eta(\Theta_A,\Theta_B,\Delta)=G(\Theta_A,\Theta_B,\Delta)+O(\eta)
\]
in \(C^2\) on a neighborhood of the reduced fold point; that is, the error and its first two derivatives are \(O(\eta)\) uniformly there.

Since the fold of \(G\) at \(\Delta_{\mathrm{fold}}^{(0)}\) is nondegenerate, standard persistence of nondegenerate saddle-node bifurcations under \(C^2\)-small perturbations implies that, for \(\eta\) sufficiently small, the full system admits a nearby fold at some
\[
\Delta_{\mathrm{fold}}=\Delta_{\mathrm{fold}}^{(0)}+O(\eta)\in(0,\pi).
\]
Thus, the full compromise branch persists for
\(
\Delta\in[0,\Delta_{\mathrm{fold}})
\)
and terminates at \(\Delta=\Delta_{\mathrm{fold}}\) through a saddle-node bifurcation.

\medskip
\noindent\textbf{Step 4: Structure of the critical mode.}
At an equilibrium of the reduced system, define
\[
\Phi_A:=\tilde\phi_A-\Theta_A,
\qquad
\Phi_B:=\tilde\phi_B-\Theta_B,
\qquad
\delta:=\Theta_B-\Theta_A,
\]
and introduce the notation
\[
A_*:=\alpha^A\cos\Phi_A,
\qquad
B_*:=\alpha^B\cos\Phi_B,
\qquad
U:=K_C\cos\delta.
\]
By the assumption that each module is on the restoring side of its leader forcing, along the relevant reduced compromise branch one has
\[
|\Phi_A|<\frac{\pi}{2},
\qquad
|\Phi_B|<\frac{\pi}{2}.
\]
This is true near the low-conflict compromise state and, by hypothesis, persists along the branch up to the fold. Hence\[
A_*>0,
\qquad
B_*>0.
\]

In this notation, the Jacobian of the reduced system~\eqref{eq:reducedR} is
\[
J_{\mathrm{red}}
=
\begin{pmatrix}
-A_*-U & U\\[2pt]
U & -B_*-U
\end{pmatrix}.
\]
At the reduced fold,
\[
\det J_{\mathrm{red}}=0,
\]
thus
\[
A_*B_*+U(A_*+B_*)=0
\]
and
\[
U=-\frac{A_*B_*}{A_*+B_*}<0.
\]
Since \(K_C>0\), this implies
\[
\cos\delta<0.
\]

Because \(\det J_{\mathrm{red}}=0\), the nullspace of \(J_{\mathrm{red}}\) is one-dimensional. A corresponding null vector is
\[
(c_A,c_B)
\propto
\bigl(U,\ A_*+U\bigr).
\]
Using the expression for \(U\), we obtain
\[
c_A \propto U<0,
\]
and
\[
c_B \propto A_*+U
=
A_*-\frac{A_*B_*}{A_*+B_*}
=
\frac{A_*^2}{A_*+B_*}>0.
\]
Therefore
\[
c_Ac_B<0.
\]

Using the reduced equilibrium parametrization
\[
\tilde{\boldsymbol\theta}
=
\Theta_A\mathbf 1_A+\Theta_B\mathbf 1_B+X(\Theta_A,\Theta_B,\Delta;\eta),
\]
we differentiate with respect to \((\Theta_A,\Theta_B)\) in the reduced null direction \((c_A,c_B)\). This gives
\[
\varphi
=
c_A\mathbf 1_A+c_B\mathbf 1_B
+
\partial_{\Theta_A}X c_A
+
\partial_{\Theta_B}X c_B.
\]
Because \(X\) is obtained from the fluctuation equations by the Implicit Function Theorem and vanishes at \(\eta=0\), it is \(O(\eta)\) in \(C^1\) on the local equilibrium manifold. Hence its derivatives with respect to \(\Theta_A\) and \(\Theta_B\) are also \(O(\eta)\), so
\[
\varphi
=
c_A\mathbf 1_A+c_B\mathbf 1_B+r,
\qquad
\|r\|=O(\eta).
\]
Moreover, \((c_A,c_B)\) lies in the kernel of the reduced Jacobian, while the fluctuation block is invertible on the zero-mean subspace. Therefore this lifted direction lies in the kernel of the full Jacobian.
Thus
\[
\varphi
=
c_A\mathbf 1_A+c_B\mathbf 1_B+O(\eta),
\qquad
c_Ac_B<0.
\]
This shows that the critical mode is an inter-module contrast mode: to leading order the two modules move coherently against one another.
\hfill\(\square\)

{\bf Remark 3.}
The non-degeneracy conditions required for the theorem have a natural interpretation as a Kuramoto-type model \cite{kuramoto1975international}. Transversality ($a \neq 0$) holds because the conflict parameter $\Delta$ enters as a direct phase shift in the leader forcing terms $-\alpha_i \sin(\theta_i-\phi_i(\Delta))$. Changing \(\Delta\) perturbs the reduced scalar equation in the critical direction, yielding the transversality coefficient \(a\neq 0\) in generic situations. Non-vanishing curvature ($b \neq 0$) occurs because the second derivative $D_\theta^2F$ involves the cosines of the phase differences $\cos(\theta_j-\theta_i)$.  Near the numerically observed loss-of-compromise point, the bottleneck phase differences are large, so the corresponding cosine terms generically do not cancel. These observations support the fold nondegeneracy conditions used in Theorem 2 and are consistent with the numerically observed loss of compromise through fold-type termination in the modular regime.



We next establish the existence of polarized decision states in the weak inter-module coupling regime and show that, when the compromise branch terminates through the fold identified in Theorem~2, these polarized states do not arise through a local bifurcation from the compromise branch. We begin with two intermediate results, and then use them to prove Theorem~3.

Let the interaction graph be partitioned into two connected modules \(M_A\) and \(M_B\), coupled by a cut-set \(C\). We introduce a small parameter \(\varepsilon\ge 0\) controlling the inter-module coupling and assume that
\[
a_{ij}\le \varepsilon
\qquad\text{for all }(i,j)\in C.
\]
We further assume that each module contains at least one informed agent, that every informed agent in \(M_A\) prefers target \(\phi_A\), and that every informed agent in \(M_B\) prefers target \(\phi_B\), with \(\phi_A\neq\phi_B\).

\subsection*{Proposition 1. (Existence of modular polarized equilibria)}
\emph{Fix a conflict angle \(\Delta=|\phi_A-\phi_B|\). For each fixed \(\Delta>0\), there exists \(\varepsilon_0(\Delta)>0\) such that, for every \(0\le\varepsilon<\varepsilon_0(\Delta)\), the full network admits a macroscopically polarized equilibrium
$\tilde{\boldsymbol\theta}^{\mathrm{pol}}(\varepsilon)
=
\bigl(
\tilde{\boldsymbol\theta}^{A}(\varepsilon),
\tilde{\boldsymbol\theta}^{B}(\varepsilon)
\bigr),$
in which the agents in \(M_A\) remain aligned near \(\phi_A\), while the agents in \(M_B\) remain aligned near \(\phi_B\). This equilibrium is locally exponentially stable.}

\medskip
\noindent\textit{Proof.}
We begin with the decoupled case \(\varepsilon=0\). Then the network splits into two independent subsystems on \(M_A\) and \(M_B\). The equilibrium of the isolated subsystem on \(M_A\) is determined only by the internal structure of \(M_A\) and the local target \(\phi_A\); it is independent of \(\phi_B\). Since all informed agents in \(M_A\) prefer the same target \(\phi_A\), this isolated subsystem has zero internal conflict. By Theorem~1 applied to \(M_A\), there exists a unique locally exponentially stable equilibrium \(\tilde{\boldsymbol\theta}^{A,0}\) near the aligned state associated with \(\phi_A\).

Similarly, the equilibrium of the isolated subsystem on \(M_B\) is determined only by the internal structure of \(M_B\) and the local target \(\phi_B\); it is independent of \(\phi_A\). Since all informed agents in \(M_B\) prefer the same target \(\phi_B\), Theorem~1 gives a unique locally exponentially stable equilibrium \(\tilde{\boldsymbol\theta}^{B,0}\) near the aligned state associated with \(\phi_B\).

Hence the product state
\[
\tilde{\boldsymbol\theta}^{\mathrm{pol}}(0)
:=
(\tilde{\boldsymbol\theta}^{A,0},\tilde{\boldsymbol\theta}^{B,0})
\]
is an equilibrium of the full decoupled system. Since the two module equilibria are centered near the distinct targets \(\phi_A\) and \(\phi_B\), the product equilibrium is macroscopically polarized across the modular cut.

The Jacobian of the full equilibrium equation at \(\tilde{\boldsymbol\theta}^{(0)}\) and \(\varepsilon=0\) is block diagonal:
\[
J^{(0)}=
\begin{pmatrix}
J_A^{(0)} & 0\\
0 & J_B^{(0)}
\end{pmatrix},
\]
where \(J_A^{(0)}\) and \(J_B^{(0)}\) are the Jacobians of the isolated module subsystems. By Theorem~1 applied to each isolated module at zero internal conflict, the Jacobians of the corresponding equilibrium equations are negative definite. Hence \(J_A^{(0)}\) and \(J_B^{(0)}\) are invertible, and therefore \(J^{(0)}\) is invertible.

Therefore, by the Implicit Function Theorem, there exists \(\varepsilon_0>0\) such that for every \(0\le\varepsilon<\varepsilon_0\), the decoupled product equilibrium persists as a smooth branch
\[
\tilde{\boldsymbol\theta}^{\mathrm{pol}}(\varepsilon).
\]
Moreover, the full dynamical linearization about the decoupled product equilibrium is locally exponentially stable at \(\varepsilon=0\). By continuity of eigenvalues, this local exponential stability persists for sufficiently small \(\varepsilon\). \hfill\(\square\)

\medskip

\subsection*{Corollary 1. (Local separation from the compromise fold)}
\emph{Assume the hypotheses of Theorem~2, so that the compromise branch \(\tilde{\boldsymbol\theta}^\ast(\Delta)\) terminates at \(\Delta=\Delta_{\mathrm{fold}}\) through a saddle-node bifurcation. Assume also that Proposition~1 applies for some \(\Delta_0>\Delta_{\mathrm{fold}}\) with \(0\le\varepsilon<\varepsilon_0(\Delta_0)\). Then the modular polarized equilibrium given by Proposition~1 does not arise through a local bifurcation from the compromise branch at \(\Delta=\Delta_{\mathrm{fold}}\).}

\medskip
\noindent\textit{Proof.}
By Theorem~2, there exists a neighborhood \(U\) of \((\tilde{\boldsymbol\theta}_c,\Delta_{\mathrm{fold}})\), where \(\tilde{\boldsymbol\theta}_c=\tilde{\boldsymbol\theta}^\ast(\Delta_{\mathrm{fold}})\), in which the equilibrium set is described by the standard fold normal form
\[
g(\xi,\mu)=a\mu+b\xi^2+\mathcal O(\mu^2,\mu\xi,\xi^3),
\qquad
\mu=\Delta-\Delta_{\mathrm{fold}},
\]
with \(a\neq 0\) and \(b\neq 0\). Standard saddle-node theory implies that, within this neighborhood, the compromise branch is the only equilibrium branch passing through the fold point, and this branch terminates at \(\mu=0\). In particular, no distinct symmetry-broken branch can emerge locally from \((\tilde{\boldsymbol\theta}_c,\Delta_{\mathrm{fold}})\).

On the other hand, Proposition~1 gives a locally exponentially stable modular polarized equilibrium for \(\Delta=\Delta_0>\Delta_{\mathrm{fold}}\) and sufficiently small \(\varepsilon\). Since no such distinct branch is created in the local fold normal form, this modular polarized equilibrium cannot arise through a local bifurcation from the compromise fold. \hfill\(\square\)

\medskip

Under the same modular weak inter-module coupling assumptions used in Proposition~1 and Corollary~1, it follows that:

\subsection*{Theorem 3. (Existence and local separation of modular polarized states)}
\emph{For each fixed conflict angle \(\Delta>0\), there exists \(\varepsilon_0(\Delta)>0\) such that, whenever the inter-module coupling satisfies \(0\le \varepsilon<\varepsilon_0(\Delta)\), there exists a macroscopically polarized equilibrium, locally exponentially stable, in which the agents in \(M_A\) remain aligned near \(\phi_A\) and the agents in \(M_B\) remain aligned near \(\phi_B\).
Moreover, if the hypotheses of Theorem~2 hold, then this modular polarized equilibrium does not arise through a local bifurcation from the compromise branch at \(\Delta=\Delta_{\mathrm{fold}}\).}

\medskip
\noindent\textit{Proof.}
The existence and local exponential stability of the modular polarized equilibrium follow directly from Proposition~1. Its local separation from the compromise branch at \(\Delta=\Delta_{\mathrm{fold}}\) follows from Corollary~1. \hfill\(\square\)

\medskip

{\bf Remark 4.}
Theorem~3 is a local statement. It shows that, in the modular weak-coupling regime, a stable modular polarized state exists, with each module aligned near its own preferred target, and that this state does not emerge from the compromise branch through a local symmetry-breaking bifurcation at \(\Delta_{\mathrm{fold}}\). It does \emph{not} by itself determine the global topology of the full equilibrium set. The broader disconnected branch geometry and the associated hysteresis picture reported in the main text are therefore supported by numerical evidence rather than by the local bifurcation argument alone.


{\bf Remark 5.}
For stable decision branches which are structurally separated from the compromise fold at $\Delta_{\mathrm{fold}}$, as observed in the numerical studies, their existence will terminate at their own distinct saddle-node limit points, which we denote $\Delta_{\mathrm{min}}$. The exact location of $\Delta_{\mathrm{min}}$ is a global property of the network geometry. In the modular continuation examples studied here, the capacity required to maintain a highly polarized state appears lower than the capacity required to maintain the global compromise state, and correspondingly the numerically observed ordering is \(\Delta_{\min}<\Delta_{\mathrm{fold}}\). When both branches remain stable across this overlap, the ordering \(\Delta_{\min}<\Delta_{\mathrm{fold}}\) produces a bistable interval
\(\Delta \in (\Delta_{\min},\Delta_{\mathrm{fold}})\), supporting hysteresis under quasi-static variations of the environmental conflict.


In Section \ref{3.3} of the main text, we describe how the collective exhibits critical slowing down just beyond the topological rupture point. When the conflict $\Delta$ slightly exceeds $\Delta_{\mathrm{fold}}$, the local compromise equilibrium is mathematically erased. However, because the vector field is strongly pinched in this regime, trajectories do not escape immediately. Instead, they experience dynamic trapping near the remnant of the destroyed equilibrium, namely the ghost of the saddle-node.
To characterize this delay and the role of uninformed damping, we first project the full second-order network dynamics onto the critical rupture mode identified in Theorem 2.

\subsection*{Proposition 2. (Local Second-Order Reduction)}
\emph{Consider the full second-order dynamics
\[
\ddot{\tilde{\boldsymbol\theta}}+\Gamma \dot{\tilde{\boldsymbol\theta}}=F(\tilde{\boldsymbol\theta},\Delta),
\]
where $\Gamma=\mathrm{diag}(\gamma_i)$ is the positive-definite damping matrix. Let $\Delta=\Delta_{\mathrm{fold}}+\mu$ for a small parameter quench $0<\mu \ll 1$. In a local neighborhood of the critical point $(\tilde{\boldsymbol\theta}_c,\Delta_{\mathrm{fold}})$, the center manifold of the first-order formulation is one-dimensional. It is convenient, however, to retain the inertial term explicitly in a lifted scalar reduction, yielding the effective second-order equation
\[
\ddot{\xi}+\Gamma_{\mathrm{eff}}\dot{\xi}
=
a\mu+b\xi^2+\mathcal{O} \left(\mu^2,\mu\xi,\xi^3,\xi\dot{\xi},\dot{\xi}^2\right),\]
where $\xi(t)$ is the scalar amplitude along the critical eigenvector \(\varphi\) from Theorem~2, with $a$ and $b$ being the saddle-node normal-form coefficients, and
\[
\Gamma_{\mathrm{eff}}:=\langle \psi,\Gamma \varphi\rangle
\]
is the effective damping of the critical mode. Because the equilibrium Jacobian is symmetric throughout the modular family considered here, the left and right critical null vectors coincide, so \(\psi=\varphi\) and therefore
\[
\Gamma_{\mathrm{eff}}=\varphi^\top\Gamma\varphi>0.
\]}

\smallbreak \textit{Proof.} Let $\Delta=\Delta_{\mathrm{fold}}+\mu$. By the center-manifold theorem for parameter-dependent dynamical systems, the invariant center manifold persists for sufficiently small $\mu>0$. At the fold, the linearized system has exactly one neutral mode, while all other perturbations decay exponentially. Thus the local dynamics are governed by a single critical direction. This retains a lifted second-order scalar description inside the critical-mode subspace because the inertial direction there is the fast direction associated with the same critical mode. In the ghost region this fast direction becomes asymptotically slaved to the slow direction, which is why the lifted second-order equation has the correct asymptotic content before the adiabatic elimination carried out in Proposition~3.

We decompose the trajectory as
\[
\tilde{\boldsymbol\theta}(t)=\tilde{\boldsymbol\theta}_c+\xi(t)\varphi+W \left(\xi(t),\dot{\xi}(t),\mu\right),
\]
where $\xi(t)\in \mathbb{R}$, the correction $W$ is orthogonal to the left eigenvector $\psi$, and $W$ contains only quadratic and higher-order terms in its arguments.

Substituting this ansatz into the governing equation yields
\[
\ddot{\xi} \varphi+\ddot{W}+\Gamma\bigl(\dot{\xi} \varphi+\dot{W}\bigr)
=
F \left(\tilde{\boldsymbol\theta}_c+\xi\varphi+W,\Delta_{\mathrm{fold}}+\mu\right).
\]
Taking the inner product with the left null vector $\psi$ eliminates the components lying in the range of the critical Jacobian $J_c$. Using the normalization $\langle \psi,\varphi\rangle=1$, we obtain
\[
\ddot{\xi}
+
\langle \psi,\Gamma\varphi\rangle \dot{\xi}
+
\langle \psi,\ddot{W}+\Gamma \dot{W}\rangle
=
\left\langle
\psi,
F \left(\tilde{\boldsymbol\theta}_c+\xi\varphi+W,\Delta_{\mathrm{fold}}+\varepsilon\right)
\right\rangle .
\]

By the standard saddle-node normal-form theorem, the projected dynamics in a neighborhood of the fold take the form
\[
g(\xi,\mu)=a\mu+b\xi^2+\cdots
\]
where $a$ and $b$ are the transversality and curvature coefficients.
Defining
\[
\Gamma_{\mathrm{eff}}:=\langle \psi,\Gamma\varphi\rangle,
\]
and observing that the projected contributions of $\dot{W}$ and $\ddot{W}$ are of higher order in the local neighborhood, we recover the stated reduced equation. \hfill$\square$

\medskip

In Proposition~3 we choose the reduced coordinate \(\xi\) and parameter offset so that \(\mu>0\) corresponds to the post-fold side and escape through the ghost proceeds in the \(+\xi\) direction. With this convention, the reduced normal form may be written with
\[ a>0, \qquad b>0.\]
Accordingly, Propositions~3 uses this fixed orientation, whereas absolute values may be inserted in invariant scaling laws when the sign convention is not specified explicitly.

Having isolated the dynamics along the critical rupture mode, we now estimate the time required for the collective to traverse the saddle-node ghost. Because the passage through the bottleneck is slow, the inertial term becomes asymptotically negligible compared to the viscous term, allowing an adiabatic elimination of the momentum.

\subsection*{Proposition 3: Asymptotic Escape Time}
\emph{In the asymptotic limit $0<\mu \ll 1$, the inertial dynamics inside the ghost region are heavily overdamped. The time $T_{\mathrm{escape}}$ required for the macroscopic trajectory to dynamically escape the bottleneck scales as
\[
T_{\mathrm{escape}}
\simeq
\frac{\pi \Gamma_{\mathrm{eff}}}{\sqrt{ ab\mu}}.
\]
Consequently, increasing the effective damping $\Gamma_{\mathrm{eff}}$ prolongs the dynamic trapping time without altering the static structural limits of the network.}

\smallbreak \textit{Proof.} In the immediate vicinity of the ghost, the vector field is $\mathcal{O}(\mu)$. We establish the dominant balance by scaling the variables. Let $\xi\sim \mathcal{O}(\sqrt{\mu})$. In order for the viscous term $\Gamma_{\mathrm{eff}}\dot{\xi}$ to balance the driving forces
\[
a\mu+b\xi^2\sim \mathcal{O}(\mu),
\]
the time derivative must satisfy $\dot{\xi}\sim \mathcal{O}(\mu)$. Consequently, the inertial term scales as
\[
\ddot{\xi}\sim \mathcal{O}(\mu^{3/2}).
\]
For $0<\mu \ll 1$, the inertial term is strictly of higher order and can therefore be adiabatically eliminated. The dynamics on the slow manifold reduce to the first-order overdamped equation
\[
\Gamma_{\mathrm{eff}}\dot{\xi}\approx a\mu+b\xi^2.
\]

The time required to traverse the bottleneck is then obtained by integrating the inverse velocity:
\[
T_{\mathrm{escape}}
\approx
\int_{-\infty}^{\infty}
\frac{\Gamma_{\mathrm{eff}}}{ a\mu+ b\xi^2} d\xi,
\]
where the infinite limits capture the leading-order asymptotic contribution of the narrow ghost region. Evaluating this standard integral yields
\[
T_{\mathrm{escape}}
\simeq
\frac{\Gamma_{\mathrm{eff}}}{\sqrt{ ab\mu}}
\left[
\arctan \left(
\sqrt{\frac{ b}{ a\mu}} \xi
\right)
\right]_{-\infty}^{\infty}
=
\frac{\pi \Gamma_{\mathrm{eff}}}{\sqrt{ ab\mu}}.
\]
This gives the stated escape-time scaling. \hfill$\square$

{\bf Remark 6.}
The effective damping
\begin{eqnarray}
\Gamma_{\mathrm{eff}}:&=& \langle \psi,\Gamma \varphi\rangle\nonumber\\
      &=& \sum_{i=1}^N \gamma_i \psi_i \varphi_i. \nonumber
\end{eqnarray}
acts as a macroscopic viscous retarder for collective rupture. Uninformed agents, characterized by $\alpha_i\approx 0$, contribute no directional bias, and therefore do not alter the static normal-form coefficients $a$ and $b$ nor the structural threshold $\Delta_{\mathrm{fold}}$.

Because $\Gamma_{\mathrm{eff}}=\varphi^\top\Gamma\varphi>0$ throughout the modular family considered here, increasing the damping of uninformed agents increases $\Gamma_{\mathrm{eff}}$ and lengthens the slowing-down episode. This is the
local dynamical mechanism underlying the long-lived trapped transients observed in Figure~\ref{fig:theorem2-simulations}(B).


In Section \ref{3.3} and Figure~\ref{fig:theorem2-simulations}(C) of the main text, we consider a slowly varying environment where the conflict is ramped continuously in time. Under finite-time observations, the macroscopic transition, namely the jump to the disconnected decision branch, is not recorded at the static topological limit $\Delta_{\mathrm{fold}}$, but at an effectively observed threshold $\hat{\Delta}_c > \Delta_{\mathrm{fold}}$.
To establish the analytical dependence of this threshold shift on the uninformed damping, we analyze the dynamic saddle-node bifurcation.

Let the conflict parameter be ramped linearly such that
\[
\Delta(t)=\Delta_{\mathrm{fold}}+vt,
\]
where $v>0$ is a small, constant sweep velocity with $0<v\ll 1$. We assume the system tracks the stable compromise branch up to $t=0$, at which point it enters the saddle-node ghost.

\subsection*{Proposition 4 (Dynamic Bifurcation Delay)}
\emph{Consider the slow continuous ramp $\Delta(t)=\Delta_{\mathrm{fold}}+vt$. In the overdamped limit, the macroscopic trajectory escapes the saddle-node ghost and jumps to the disconnected branch at an effective conflict threshold $\hat{\Delta}_c$. In the asymptotic limit $v\to 0$, the dynamic threshold shift scales as
\[
\hat{\Delta}_c-\Delta_{\mathrm{fold}}\propto (\Gamma_{\mathrm{eff}}v)^{2/3},
\]
where $\Gamma_{\mathrm{eff}}$ is the effective damping of the critical mode. Consequently, increasing the damping of uninformed participants monotonically shifts the observed transition to higher apparent conflict levels.}

\smallbreak \textit{Proof.} We begin with the one-dimensional reduced dynamics established in Proposition~2. Because the transition through the bottleneck is slow, the inertial term $\ddot{\xi}$ is negligible in the ghost region. Applying the adiabatic elimination used in Proposition~3, the dynamics within the ghost region are governed by the overdamped Riccati equation
\[
\Gamma_{\mathrm{eff}}\dot{\xi}\approx a\bigl(\Delta(t)-\Delta_{\mathrm{fold}}\bigr)+b\xi^2.
\]
Substituting the linear ramp $\Delta(t)-\Delta_{\mathrm{fold}}=vt$, we obtain
\[
\Gamma_{\mathrm{eff}}\frac{d\xi}{dt}=avt+b\xi^2,
\]
where $a>0$ and $b>0$ are the static normal-form coefficients evaluated at the fold.

\begin{figure*}[t]
\includegraphics[width=12cm]{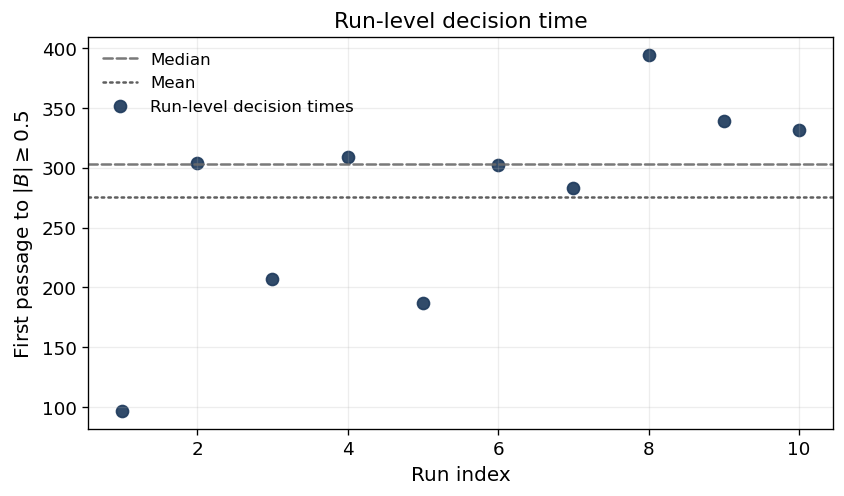}
\caption{\textbf{Run-level decision times in the Kilobot experiments.}
Each point denotes the first-passage time at which an individual run satisfies $|B(t)|\ge 0.5$. The dashed and dotted horizontal lines indicate the median and mean decision times, respectively. The concentration of run-level decision times around $t\approx 300$ supports the interpretation of Fig.~\ref{fig:kilobots_bias}(B), where the inter-run variance reaches its maximum at nearly the same time.}
\label{fig:kilobots_firstpassage_si}
\end{figure*}
To determine the dominant balance and the characteristic escape time, we non-dimensionalize the equation. We introduce the scaling transformations
\[
t=\tau s,\qquad \xi=\beta u,
\]
where $\tau$ is the characteristic time scale of the delay, $s$ is dimensionless time, and $u$ is dimensionless amplitude. The transformed equation is
\[
\frac{\Gamma_{\mathrm{eff}}\beta}{\tau}\frac{du}{ds}=av\tau s+b\beta^2u^2.
\]
To balance the viscous drag, the driving force, and the nonlinear curvature equally, we require the coefficients of all three terms to be of the same order:
\[
\frac{\Gamma_{\mathrm{eff}}\beta}{\tau}\sim av\tau,
\qquad
\frac{\Gamma_{\mathrm{eff}}\beta}{\tau}\sim b\beta^2.
\]
Solving this system for the scaling factors $\tau$ and $\beta$ yields
\[
\beta\sim \frac{\Gamma_{\mathrm{eff}}}{b\tau}.
\]
Substituting $\beta$ into the first balance condition gives
\[
\frac{\Gamma_{\mathrm{eff}}^2}{b\tau^2}\sim av\tau
\qquad\Longrightarrow\qquad
\tau^3\sim \frac{\Gamma_{\mathrm{eff}}^2}{abv}.
\]
Thus, the characteristic time spent traversing the dynamic bottleneck scales as
\[
\tau\propto \Gamma_{\mathrm{eff}}^{2/3}(abv)^{-1/3}.
\]

The macroscopic jump occurs at a dimensionless time $s_c\sim \mathcal{O}(1)$, corresponding to the physical escape time $t_{\mathrm{escape}}\propto \tau$. The effectively observed transition threshold is defined as the conflict level at the moment of escape:
\[
\hat{\Delta}_c=\Delta_{\mathrm{fold}}+vt_{\mathrm{escape}}.
\]
Substituting the time scale $\tau$, the dynamic threshold shift is
\[
\hat{\Delta}_c-\Delta_{\mathrm{fold}}
\propto
v\left(\Gamma_{\mathrm{eff}}^{2/3}a^{-1/3}b^{-1/3}v^{-1/3}\right)
=
\mathcal{C}(\Gamma_{\mathrm{eff}}v)^{2/3},
\]
where $\mathcal{C}=(ab)^{-1/3}$ is a structural constant depending only on the static fold geometry.

This establishes the classical $\mathcal{O}(v^{2/3})$ dynamic bifurcation-delay scaling. Crucially, the shift is proportional to $\Gamma_{\mathrm{eff}}^{2/3}$. Because uninformed agents contribute direction-free dissipation ($\gamma_i>0$) to the critical rupture mode, their presence increases $\Gamma_{\mathrm{eff}}$. Therefore, the addition of passive participants generates a monotonic apparent increase in the observed structural capacity under finite-time ramping, recovering the continuous threshold shifts observed in Figure~\ref{fig:theorem2-simulations}(C). \hfill$\square$

\bigskip

\section*{Supporting analysis for the Kilobot swarm experiments}
To complement Fig.~\ref{fig:kilobots_bias} in the main text, Fig.~\ref{fig:kilobots_firstpassage_si} reports the distribution of run-level decision times across the 10 independent Kilobot runs. Here, the decision time is defined as the first-passage time at which the macroscopic bias satisfies $|B(t)|\ge 0.5$. This auxiliary analysis supports the observation that the peak in inter-run variance occurs at nearly the same time as a typical run first reaches a macroscopically polarized state.


\end{document}